\documentclass[12pt]{article}
\usepackage{epsf}
\usepackage{amsmath}
\usepackage{amssymb}
\newcommand{\CFT}{\mathrm{CFT}}
\newcommand{\be}{\begin{equation}}
\newcommand{\bea}{\begin{eqnarray}}
\newcommand{\ena}{\end{eqnarray}}
\newcommand{\non}{\nonumber\\ }
\newcommand{\en}{\end{equation}}
\newcommand{\beq}{\begin{equation}}
\newcommand{\eeq}{\end{equation}}
\newcommand{\beqn}{\begin{eqnarray}}
\newcommand{\eeqn}{\end{eqnarray}}
\newcommand{\dd}{{\mathrm d}}
\newcommand{\lan}{{\langle}}
\newcommand{\ran}{{\rangle}}

\begin{document}
\hfill{Landau-TMP 01/03}

\vspace{3.0cm}

\begin{center}

{\Large
On correlation functions
in the perturbed \\
\vspace{1mm}
minimal models ${{\cal M}}_{2,2n+1}$
}

\vskip 1.0cm {\large
A.~A.~Belavin${}^{a}$, V.~A.~Belavin${}^{b}$, A.~V.~Litvinov${}^{a}$,\\
Y.~P.~Pugai${}^{a}$ and Al.~B.~Zamolodchikov${}^{c}$}\\

\vspace{.4cm}
{ \it
$^a$ L.D.~Landau Institute for Theoretical Physics,\\
Chernogolovka 142432, Russia

\vspace{0.3cm}
$^b$ ITEP, B. Cheremushkinskaya 25, Moscow, 117259, Russia

\vspace{0.3cm}
$^c$ Laboratoire de Physique Mathematique, University Montpellier II,\\
Pl.E.Bataillon, 34095 Montpellier, France
}

\end{center}


\date{}


\begin{abstract}
Two-point correlation functions of spin operators
in the minimal models ${{\cal M}}_{p,p'}$ perturbed by the field $\Phi_{13}$
are studied in the framework of conformal perturbation theory~\cite{AlZamLY}.
The first-order corrections for the structure functions are
derived analytically in terms of gamma functions. Together
with the exact vacuum expectation values of local operators, this gives the
short-distance expansion of the correlation functions.

The long-distance behaviors of these
correlation functions in the case ${{\cal M}}_{2,2n+1}$ have been worked out
using a form-factor bootstrap approach.

The results of numerical calculations demonstrate that the short- and
long-distance expansions match at the intermediate distances.
Including the descendent operators in the OPE drastically improves the
convergency region.
The combination of the two methods thus describes the correlation
functions at all length scales with good precision.
\end{abstract}

\section{Introduction}
A complete set of correlation functions is
the main object completely characterizing field theories.
The well-known nontrivial examples where exact correlation functions have been
found are the two-dimensional scaling Ising model in the zero
magnetic field and the conformal field theories~\cite{BPZ,McCoySato}.
Unfortunately, away from the free fermion/conformal points,
the specific methods that have been applied for those
models are difficult to generalize.
In particular, the problem of finding
exact analytic expressions for
correlation functions of integrable
two dimensional massive models is still open.
In approaching this problem, we follow~\cite{AlZamLY} and study
general behaviors of
correlation functions in the perturbed conformal field theories at all scales
by applying a combination of conformal perturbation theory and the
form-factor bootstrap approach.

In what follows, we concentrate on the minimal models
of CFT ${{\cal M}}_{p,p'}$ (with coprime integers
$1<p<p'$) perturbed by the field $\Phi_{13}$.
These massive field theories having an infinite number of
conservation laws are integrable~\cite{Zam}.
From the statistical mechanics standpoint, the given massive
models describe a universality class of the integrable RSOS-type
models~\cite{ABF} in the corresponding antiferromagnetic regime.

The conformal field theories ${{\cal M}}_{p,p'}$ underlying the massive models
under consideration are characterized by the central charge
of the Virasoro algebra
\be
\label{CenChaCFT}
{
c=1-6\frac{(p'-p)^2}{pp'}\,.
}
\end{equation}
There are $(p-1)\times (p'-1)/2$ primary fields $\Phi_{l,k}$
($l=1,\ldots, p-1$ and $k=1,\ldots,p'-1$)
in the model. The conformal dimensions $(\Delta_{l,k},\Delta_{l,k})$ of
these fields are determined by the Kac formula:
\be
\label{SpectCFT}
{
\Delta_{l,k}=\frac{(p'l-p k)^2-(p'-p)^2}{4pp'}\,.
}
\end{equation}
In what follows, we impose the standard normalization for the primary fields
adopted in CFT,
\be
\label{Normal}
\langle \Phi_{l,k}(x)\Phi_{l,k}(0)\rangle_{\CFT}={|x|^{-4\Delta_{l,k}}}\,.
\end{equation}
We use the symbol $I=\Phi_{11}$ for the unity operator throughout this work.
The operator $\Psi=\Phi_{12}$ is identified with the spin operator,
and $\Phi=\Phi_{13}$, with the energy operator.
It is convenient to use the positive rational parameter
$$
\xi=\frac{p}{p'-p}
$$
instead of $p$ and $p'$ because, for example, the operators
$I$, $\Psi$, $\Phi$, and $\Phi_{15}$, which are important for us,
then have the conformal dimensions
$$
\Delta_I=0\,,\quad \Delta_\Psi=\frac{\xi-2}{4(\xi+1)}\,,\quad \Delta_\Phi=\frac{\xi-1}{\xi+1}\,,
\quad \Delta_{\Phi_{15}}=\frac{4\xi-2}{\xi+1}\,.
$$

The general procedure developed in~\cite{BPZ,DotFat}
allows computing structure constants in the operator algebra
and determining all correlation functions. This knowledge
can be used to study correlation functions of the
$\Phi_{13}$ perturbation of the
model ${{\cal M}}_{p,p'}$ as follows.

The scaling model we study can be formally defined
through the action
\bea
\label{ActiRSG}
{{\cal A}}_{{{\cal S}}_{p,p'}}={{\cal A}}_{{{\cal M}}_{p,p'}}+g \int \Phi\ d^2 x\ ,
\ena
where ${{\cal A}}_{{{\cal M}}_{p,p'}}$ denotes
the ``action" of the critical model and
$g$ is a coupling constant having the dimension
$g\sim m_1^{2-2\Delta_{\Phi}}$, where $m_1$ is a mass gap
(the mass of lightest particle).
We note that
the operator $\Phi$ has the conformal dimension
$\Delta_\Phi<1$ and the perturbation is relevant. More explicitly,
the dimensional parameters $g$ and $m_1$ are connected via the exact
``mass--coupling-constant relation" found in~\cite{AlZamMassMu}:
\be
\label{MassCouplRSG}
{
\pi g=-\ \frac{(\xi+1)^2}{(\xi-1)(2\xi-1)}
\Bigl(\gamma(\frac{3\xi}{\xi+1})\gamma(\frac{\xi}{\xi+1}) \Bigr)^{\frac{1}{2}}
m^{\frac{4}{\xi+1}}\,.
}
\end{equation}
Here and hereafter, the massive parameter $m$ is related to the
mass of the lightest particle $m_1$ by
$$
m=m_1\left(\frac{\pi\gamma(1-\frac{\xi}{2})
\gamma(\frac{1+\xi}{2})}{8\sin\pi\xi}\right)^{\frac{1}{2}}\,
$$
and we set
$$
\gamma(x)\;=\;\frac{\Gamma(x)}{\Gamma(1-x)}\,.
$$

Formal definition~\eqref{ActiRSG} is understood in the sense of
the perturbation series. For example, for the correlation function
of the spin fields, we have
$$
\langle \Psi(x)\Psi(0)\rangle =\sum_{n=0}^{\infty}\frac{(-g)^n}{n!}\int\langle \Psi(x)\Psi(0)
\Phi(y_1)\cdots\Phi(y_n)\rangle_{\CFT}d^2y_1\cdots d^2y_n\,.
$$
Here the symbols $\Psi$ in the l.h.s.\ denote the scaling fields
in the perturbed model; in the UV limit, these fields become
the conformal spin fields $\Phi_{12}$. Throughout this work,
we generally assume that the renormalized fields in~\eqref{ActiRSG} have definite
scaling dimensions and are denoted by the same letter as
in the conformal $g\to 0$ limit. We also require that the normalization
of scaling fields at the limit be fixed as in~\eqref{Normal}.

The UV and IR regularization schemes
for the integrals over the plane were discussed extensively in~\cite{AlZamLY}.
We only briefly recall the final prescription here. Since the
perturbing field is relevant, the UV renormalization
can be achieved by adding a finite number of counterterms.
To handle the IR divergences and to
work out the infrared-safe perturbation theory, the so-called
conformal perturbation theory is
developed~\cite{AlZamLY}.\footnote{See~\cite{GuiMag} for the
conformal perturbation theory in all orders.}
The conformal perturbation theory is based on the following hypothesis.
Let $\{A_{a},\ a=0,1,\ldots\}$ be a complete set of
local fields in the scaling theory.
It is assumed that if the renormalized fields that are eigenvectors of the
dilatation operator, i.e., the fields with a definite
scaling dimensions, are chosen as basis elements,
then the structure functions in the corresponding operator-product expansion
\be
\label{ScaOPE}
A_m(x)A_{a}(0)=\sum_{b} C_{A_mA_a}^{A_b}(r)A_{b}(0)\,,
\end{equation}
are analytic functions of the coupling constant $g$.
This is a rather natural conjecture since the functions
$C_{A_mA_a}^{A_b}(r)$ as local quantities are assumed to not develop
any nonanalyticity.
We propose that under this choice, the renormalized fields
$A_b(r)$ turn out to be
perturbations of the
corresponding basis fields (primaries and descendants)
of the conformal model.
Then the functions $C_{A_mA_a}^{A_b}(r)$
are given by regular expansions
\be
\label{StruFun}
C_{A_mA_a}^{A_b}(r)= r^{2(\Delta_{b}-\Delta_{a}-\Delta_{m})}
\Bigl(C_{A_mA_a}^{A_b}+gr^{2-2\Delta}Q^{(1)}+
(gr^{2-2\Delta})^2Q^{(2)}+\cdots\Bigr)\,.
\end{equation}
The zeroth-order terms here are determined by the structure
constants $C_{A_mA_a}^{A_b}$ from the critical model,\footnote{We note that
because of fusion rules, many of the $C_{A_mA_a}^{A_b}$ in the
conformal case are zeroes. But some of the
corresponding $C_{A_mA_a}^{A_b}(r)$ may become
nonzero because of the correction terms.}
and the problem of perturbation theory
is to define $g$-independent corrections $Q^{(i)}$.

To define the correlation functions following~\eqref{ScaOPE},
we must also find vacuum expectation values (VEVs)
of the local fields $A_a$, which might be nonzero in the off-critical case.
These important quantities are nonlocal and, generally, nonanalytic in
the coupling constant $g$.
From counting dimensions, we have
$
\lan A_b\ran\sim m^{2\Delta_{b}}\,.
$
The set of VEVs contain all nonperturbative
information on the theory. The VEVs for perturbed primaries were
proposed in~\cite{LZ} (also see~\cite{FLZZ}).
The Lukyanov and Zamolodchikov formula for the VEVs in the first ground
state~\cite{ABF} can be written as
\bea \label{VEVPrimCF} &&\lan\Phi_{1k}\ran=(-1)^{k-1}
m^{2\Delta_{1k}}Q(1-\xi(k-1))\,. \ena The function $Q(\eta)$ is
defined in terms of the following exponent of an integral that is
understood in the sense of analytic continuation from the region
where it converges:
\bea \label{Qfactor}
Q(\eta)=\exp \int_{0}^\infty \frac{dt}{t}
\Bigl(\frac{\cosh(2t)\sinh((\eta-1)t)\sinh((\eta+1)t)}
{2\cosh(t)\sinh(\xi t)\sinh((\xi+1)t)}
-\frac{\eta^2-1}{2\xi(\xi+1)}e^{-2(\xi+1)t} \Bigr)\,. \ena
%
We note that if $\xi>1$, for instance, as in the models
${{\cal M}}_{p,p+1}$ of principal series perturbed by the energy operator,
then there are no bound states.
In that case, we use the kink mass in the
expressions like~\eqref{VEVPrimCF}.

The short distance expansion for the two-point correlation
function of spin fields is determined by the VEVs of fields
$\Phi_{1k}$ (and their descendants) with odd $k=2s+1$ (see Eq.
(\ref{TwoI}) below). After some computations we found that for
these fields the expression (\ref{Qfactor}) simplifies and can be
written explicitly in terms of gamma functions as following
\begin{equation}\label{QgammaI}
Q(1-2s\xi)=\prod_{l=1}^{s}\mathcal{Q}_l(\xi)\,,
\end{equation}
where
\begin{multline}\label{QgammaII}
   \mathcal{Q}_l(\xi)=\frac{(1-(2l-1)\xi)(1-2l\xi)}{2(\xi+1)}\times\\
   \times
   \gamma\left(\frac{(2l-1)\xi}{2}\right)
   \gamma\left(\frac{1-(2l-1)\xi}{2}\right)
   \left[
   \gamma\left(\frac{1-2(l-1)\xi}{\xi+1}\right)
   \gamma\left(1-\frac{(2l+1)\xi}{\xi+1}\right)
   \right]^{\frac{1}{2}}.
\end{multline}

The one-point VEVs for the first nontrivial descendent operators
are also known in the analytic form~\cite{FFLZZ},
\bea
\label{VEVDes}
&&\lan L_{-2}\bar{L}_{-2}\Phi_{1k}\ran=-(1+\xi)^4
{{\cal W}}(1-\xi(k-1))
m^4\lan\Phi_{1k}\ran \,,
\ena
where the function ${{\cal W}}(\eta)$ is
$$
{{\cal W}}(\eta):=\frac{1}{\xi^2(\xi+1)^2}
\gamma(\frac{1+\eta+\xi}{2})\gamma(\frac{\eta-\xi}{2})
\gamma(\frac{1-\eta+\xi}{2})\gamma(-\frac{\eta+\xi}{2})\,.
$$

Although the problem of determining exact VEVs of all descendants
is still open, the knowledge achieved up to now can already be used
to analyze correlation functions.
For example, VEVs~\eqref{VEVPrimCF}--\eqref{VEVDes} determine
the leading terms in the short-distance expansion of the
two-point correlation function of spin operators,
\begin{eqnarray}
\label{TwoI}
&&\lan\Psi(x)\Psi(0)\ran= C^{I}_{\Psi\Psi}(r)\lan I \ran + C^{\Phi}_{\Psi\Psi}(r)\lan\Phi(0)\ran
+C^{\Phi_{15}}_{\Psi\Psi}(r)\lan \Phi_{15} \ran
\non
&&\hspace{0.5cm} +\ C^{L_{-2}\bar{L}_{-2}I}_{\Psi\Psi}(r)
\lan L_{-2}\bar{L}_{-2}I \ran +C^{L_{-2}\bar{L}_{-2}\Phi}_{\Psi\Psi}(r)\lan L_{-2}\bar{L}_{-2}\Phi(0)\ran+\cdots\,,
\end{eqnarray}
which is the main object of study in the first part of this work.
The conformal perturbation theory based on an exact knowledge of VEVs of
local operators is thus applicable for an effective study of the
short-distance ($mr\ll1$) behaviors of correlation functions.

On the other hand, theories~\eqref{ActiRSG} are massive integrable
models with factorized scattering.
The scattering matrices in those
theories coincide with those in the
restricted sine-Gordon model~\cite{SmirnovQG,LeClair,RSOS}.
We can apply the spectral
expansion~\cite{Karowski,Smirnov} (see~\cite{Mussar} and
the references therein for details)
\begin{eqnarray}
\label{TwoII}
&&\lan\Psi(x)\Psi(0)\ran=\sum_{n=0}^{\infty}\sum_{\{a_n\}}\frac{1}{n!}\int
\frac{d \beta_1}{2\pi}\cdots \frac{d \beta_n}{2\pi}
\lan 0|\Psi(0)|\beta_1,\ldots,\beta_n\ran_{a_1\cdots a_n} \times\non
&&\hspace{2.0cm} \times\ {}_{a_1\cdots a_n}{}\lan \beta_1,\ldots,\beta_n)
|\Psi(0)|0\ran
\ e^{-r\sum m_j\cosh\beta_j }
\end{eqnarray}
for the correlation functions and study the same correlation functions
at long distances $mr\gg1$. Here, $a_i$ are the particle types,
and the quantities
$$
\lan 0|\Psi|\beta_1,\ldots,\beta_n)\ran_{a_1\cdots a_n}\,, \hspace{0.5cm}
{}_{a_1\cdots a_n}{}\lan \beta_1,\ldots,\beta_n)|\Psi|0\ran
$$
are matrix elements of the local operators in the basis
of asymptotic states, the form factors.
The latter are assumed to be meromorphic
functions of the rapidity variables $\beta$ satisfying
a set of locality axioms~\cite{Smirnov}. For the case of primary fields,
the exact solutions for the form-factor equations
have already been found~\cite{AlZamLY,Smirnov,SmirnovQG,Koubek,MussarI,LuSine}.

The difficulty not easily handled
in the form-factor approach is that the axioms for
scalar primary fields do not fix
an overall normalization of the
form factors or, in other words, the normalization of
local operators. To impose conformal normalization~\eqref{Normal},
we determine the overall rapidity-independent factor
as in~\cite{AlZamLY,LuSine} by
the VEVs of local operators~\eqref{VEVPrimCF}
(also see~\cite{VEVAlg}). This prescription seems quite natural
from the standpoint of the
quantum group procedure~\cite{AlZamMassMu,SmirnovQG,FLZZ}.

In the integrable cases, there are thus two different
expansions~\eqref{TwoI} and~\eqref{TwoII} of the
same correlation functions of the local operators. A
natural idea would be to compare these formulae for short
and long distances. As soon as the expansions are matched somewhere at
the intermediate distances, we would obtain
a description of correlation functions at all scales.
The practical question, of course, is how easily that can be achieved.

We mention that this program of studying correlation functions,
initiated in~\cite{AlZamLY} (also see~\cite{FFLZZ}) for the case of the
Lee--Yang model, has already been applied to several more complicated
cases~\cite{MagnoliII,CFTPerturb,LukDoy}.

\begin{figure}[!htb]
\vspace{2mm}
\begin{center}
\epsfxsize=10.0cm \epsffile{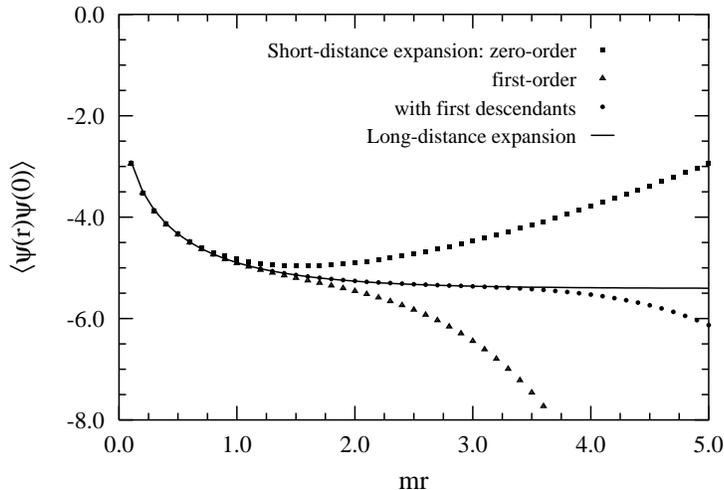}
\end{center}
\caption{Correlation function of the two spin operators (in units
$m^{\frac{8}{7}}$) in the minimal model ${{\cal M}}_{2,7}$
perturbed by the field $\Phi_{13}$.
}
\label{corrfun1}
\end{figure}

We have found that for the models ${{\cal M}}_{p,p'}$
perturbed by the field $\Phi_{13}$ in the resonance-free cases,
an elegant analytic form for the first-order perturbation theory
for the structure functions can be easily provided.
With answers~\eqref{VEVPrimCF}--\eqref{VEVDes} taken into account,
this leads to expression~\eqref{AnswerShort} controlling
the UV behaviors of the spin correlation functions.

For the case ${{\cal M}}_{2,2n+1}$, this together
with the form-factor expansion up to
three particles already gives a broad region where the two expansions
match.\footnote{We choose the model ${{\cal M}}_{2,2n+1}$ just for simplicity.
The form factors of primaries in this case do not contain contour integrals.
Another perturbed CFT with that property is a parafermionic model perturbed by
the first energy field~\cite{ABFII}.}
For example, one of the plots illustrating this phenomenon
for the model ${{\cal M}}_{2,7}$ perturbed by the energy
operator is shown in Fig.~\ref{corrfun1}.
The large region of convergence between 1 and 3 makes it clear that
the given method provides a good approximation
for correlation functions at all scales.

This matching between different
expansions at the very
first steps demonstrates that the combination
of the conformal-perturbation and form-factor
approaches is a very useful tool for
checking the consistency of different
proposals used in both constructions.
For instance, in our case, we obtain an additional
confirmation for statements
that the form factors for the primary fields are given
by the minimal solutions with the prescribed analytic
properties~\cite{Koubek,LuSine}, that the normalization of the
form factors is fixed correctly, and so on. Moreover,
that the contribution from the descendent operators
drastically improves the convergence of the two expansions
(see Fig.~\ref{corrfun1}) can be treated
as a good check for the correctness of~\eqref{VEVDes}.
In addition, this method can
also be treated as an alternative way for
studying universal scaling behaviors of the correlation
functions in the corresponding lattice RSOS models~\cite{ABF,ABFIII}.

Although many technical components that we used are already known,
we collect all necessary facts for the reader's convenience and
to make the text self-consistent.
This work is organized as follows.
In Sec.~2, we consider the procedure
for finding first-order corrections
to the structure functions.
We demonstrate that the double integrals over a complex plane
in the physically interesting situations can be taken
analytically leading to an elegant answer in terms of gamma functions.
We thus obtain the UV expansion for
the correlation function of spin operators.
We provide some details of the numerical computation for
${{\cal M}}_{2,7}$ as an example.
In Sec.~3, we recall the scattering data and
form-factor axioms applicable to our case with many
particles. We restrict our
attention to the three-particle case, which already gives
good IR expansion formulae. In Sec.~4, we
give some results of the numerical
computations and demonstrate the
consistency of both expansions. In Sec.~5, we draw final conclusions.
The technical details are collected in the appendix.

\section{Short-distance expansion scaling
${{\cal M}}_{p,p'}$ model}
Our main aim is to study two-point correlation functions
of the spin operators in the
massive $\Phi_{13}$ perturbation of ${{\cal M}}_{p,p'}$
model~\eqref{ActiRSG}. In this section, we analyze the
short-distance expansion based on the operator algebra
\bea
\label{TwoPRSG}
&&\lan\Psi(x)\Psi(0)\ran= C^{I}_{\Psi\Psi}(r)\lan I \ran + C^{\Phi}_{\Psi\Psi}(r)\lan\Phi(0)\ran
+C^{\Phi_{15}}_{\Psi\Psi}(r)\lan \Phi_{15} \ran
\non
&&\hspace{0.5cm} +\ C^{L_{-2}\bar{L}_{-2}I}_{\Psi\Psi}(r)
\lan L_{-2}\bar{L}_{-2}I \ran +C^{L_{-2}\bar{L}_{-2}\Phi}_{\Psi\Psi}(r)\lan
L_{-2}\bar{L}_{-2}\Phi(0)\ran+\cdots\,.
\ena
In the r.h.s., we keep only the leading terms in the small $r$
that have nonzero VEVs. We thus omit the derivative operators
as well as operators with nonzero spins. Taking the known VEVs
of local operators~\eqref{VEVPrimCF}--\eqref{VEVDes},
we study correlation functions~\eqref{TwoPRSG}
following the program initiated in~\cite{AlZamLY}.

\subsection{Perturbation theory for structure functions}
According to~\cite{AlZamLY}, the structure functions
in~\eqref{TwoPRSG} have the expansion with the zeroth order given by
the structure constants $C_{\Psi\Psi}^{K}$
from the underlying ${{\cal M}}_{p,p'}$~\eqref{StruFun},
\be
\label{AlZamPreIRSG}
C_{\Psi\Psi}^{K}(r)=C_{\Psi\Psi}^{K}\ r^{2\Delta_K -4\Delta_\Psi}
+{C}^{K\ (1)}_{\Psi\Psi}(r)+\cdots\,.
\end{equation}
The first-order corrections ${C}^{K\ (1)}_{\Psi\Psi}(r)$ can
be represented in the regularized form with $R$
being the parameter of the infrared cutoff,
\bea
\label{FirstCorrCorr}
&&{C}^{K\ (1)}_{\Psi\Psi}(r):=\lim_{R\to \infty}\Bigl[
-g \int_{|y|<R}\lan\tilde{{{\cal A}}}^{K}(\infty)\Phi(y)\Psi(x)\Psi(0)\ran_{\CFT}d^2y \non
&&\hspace{2.3cm}+\pi g \sum_{A}\frac{C^{A}_{\Psi\Psi}C^{K}_{\Phi A}}
{\Delta_K-\Delta_A-\Delta_\Phi+1} R^{2(\Delta_K-\Delta_A-\Delta_\Phi+1)}r^{2\Delta_A-4\Delta_\Psi}
\Bigr]\,.
\ena
As in the Lee--Yang case~\cite{AlZamLY},
it is convenient for practical purposes to apply an analytic
regularization procedure for the integral.
In our case of two operators $\Psi\equiv \Phi_{12}$,
we analyze the regularized integrals
\bea
\label{FirsStruRSG}
&& {C}^{I(1)}_{\Psi\Psi}(r)=-g \int'\lan\Phi(y)\Psi(x)\Psi(0)\ran_{\CFT}d^2y\,, \non
&&{C}^{\Phi(1)}_{\Psi\Psi}(r)=-g \int'\lan\Phi(\infty)\Phi(y)\Psi(x)\Psi(0)\ran_{\CFT}d^2y\,,\\
&& {C}^{\Phi_{15}(1)}_{\Psi\Psi}(r)=-g \int'\lan\Phi_{15}(\infty)\Phi(y)\Psi(x)\Psi(0)\ran_{\CFT}d^2y\ .\nonumber
\ena
We recall the exact form of the conformal correlators appearing here.
The correlation functions in the integral in the first line above
are defined via the structure constants of the operator algebra
\be
\label{ThreeCFT}
\lan\Phi(y)\Psi(x)\Psi(0)\ran_{\CFT}=C_{\Psi\Psi}^{\Phi}|y|^{-2\Delta_\Phi}|x|^{-2(2\Delta_\Psi-\Delta_\Phi)}
|y-x|^{-2\Delta_\Phi}\,.
\end{equation}
The correlation function including the
field $\Phi_{15}$ has a very similar form. Indeed,
the correction term in the third line can be rewritten as
\begin{eqnarray*}
&&{C}^{\Phi_{15}(1)}_{\Psi\Psi}(r)=
-g r^{2(1-\Delta_\Phi-2\Delta_\Psi+\Delta_{15})} \non
&&\hspace{2.2cm}\times
\int'|z|^{2(\Delta_\Phi+2\Delta_\Psi-2-\Delta_{15})}
\lan\Phi_{15}(\infty)\Phi(1)\Psi(z)\Psi(0)\ran_{\CFT}d^2z\,,
\end{eqnarray*}
where
\be
\lan\Phi_{15}(\infty)\Phi(1)\Psi(z)\Psi(0)\ran_{\CFT}=C_{\Psi\Psi}^{\Phi}C_{\Phi\Phi}^{\Phi_{15}}
\ |z|^{\frac{\xi}{\xi+1}}\ |1-z|^{\frac{2\xi}{\xi+1}}\,.
\end{equation}
In general, the four-point correlation functions
of the primary fields in the conformal
models can be found, for example, using the free-field construction of
Dotsenko and Fateev~\cite{DotFat} or the differential equations method.
For example, the integral in the second
line of Eq.~\eqref{FirsStruRSG} is expressed
through the correlation function
\bea
\label{ConfCorr}
&&\langle \Psi(0)\Psi(z)\Phi(1)\Phi(\infty)\rangle_{\CFT}\non
&&\quad=
C_{\Psi\Psi}^{\Phi}C_{\Phi\Phi}^{\Phi}\ \Bigl| z^{\frac{\xi}{4(\xi+1)}}\ (1-z)^{\frac{\xi}{2(\xi+1)}}
\ _2F_1\Bigl(\frac{\xi}{\xi+1}, \frac{3\xi-1}{\xi+1},\frac{2\xi}{\xi+1}|z\Bigr) \Bigr|^2
\non
&&\quad \hspace{1.6cm}+\ \Bigl| z^{\frac{2-\xi}{4(\xi+1)}}\ (1-z)^{\frac{\xi}{2(\xi+1)}}
\ _2F_1\Bigl(\frac{2}{\xi+1}, \ \frac{2\xi}{\xi+1},\ \frac{2}{\xi+1}|z\Bigr) \Bigr|^2\,.
\ena
With Eq.~\eqref{BB}, it has the form
\begin{eqnarray}
\label{CPPPInt}
&&
{C}^{\Phi(1)}_{\Psi\Psi}(r)=-g r^{2-4\Delta_\Psi}
\int'|z|^{2\Delta_\Psi-2}\langle \Psi(0)\Psi(z)\Phi(1)\Phi(\infty)\rangle_{\CFT}d^2z\non
&&\hspace{1.6cm}
=-g r^{\frac{\xi+4}{\xi+1}}\frac{\gamma(\frac{2}{\xi+1})}{\gamma^2(\frac{1}{\xi+1})}
\lim_{\epsilon\to 0} \biggl(\int \dd^2x \int \dd^2y \ |x|^{-\frac{2\xi}{\xi+1}} |1-x|^{-\frac{4\xi}{\xi+1}}
\non
&&\hspace{5.7cm}\times
|y|^{-2\frac{\xi+3}{\xi+1}+4\epsilon}|1-y|^{\frac{2\xi}{\xi+1}-2\epsilon} |x-y|^{-\frac{2\xi}{\xi+1}}
\biggr) \,.
\end{eqnarray}
Here, we fixed the way of an
analytic regularization in the second line that allows computing the
integral explicitly before taking the limit. To do this, we use
formula~\eqref{IntegrGamma}, which is derived in the next section.

\subsection{The integrals}
From explicit formulae~\eqref{ThreeCFT}--\eqref{ConfCorr},
we find that the properly regularized corrections~\eqref{FirsStruRSG}
are reduced to integrals of the forms
\bea
\label{BeIIntRSG}
J_1(p,q)&=&\int d^2 y \ |y|^{2p}|y-1|^{2q}\,,\\
%
\label{DoublNumARSG}
J_2(a,b,d,e,c)
&=&
\int \dd^2x \int \dd^2y \ |x|^{2a} |1-x|^{2b} |y|^{2d}|1-y|^{2e} |x-y|^{2c} \,.
\ena
The first integral can be easily computed by reducing it to
a product of beta-integrals, giving
\be
\label{BeIntGaRSG}
{
J_1(p,q)=
\frac{\pi \gamma(p+1)\gamma(q+1)}{\gamma(p+q+2)}\,.
}
\end{equation}

Integral~\eqref{DoublNumARSG} is more complicated.
It can be rewritten~\cite{Lipatov,DotFat,ConstFlum}
in terms of a product of two contour integrals
admitting a representation via the higher
hypergeometric functions $_3F_2$ at unity~\cite{DotPP}
(also see~\cite{CFTPerturb}).
For completeness, we collect the details of computations
in Appendix B.

In our cases~\eqref{FirstCorrCorr}, the integral
$J_2(a,b,d,e,c)$ appears at special values of the parameters
and can be expressed in terms of gamma functions
due to the following formula
\bea
\label{IntegrGamma}
&&J_2\biggl(\frac{\beta-\alpha-1}{2},\delta-\beta-1,\alpha-1,-\frac{\delta}{2},
\frac{\beta-\alpha-1}{2}\biggr)
\non
&&\hspace{1.5cm}= 2^{2\alpha-2}\pi^2 \gamma\biggl(\frac{\alpha}{2}\biggr)
\gamma\biggl(\frac{\beta}{2}\biggr)\gamma\biggl(\frac{1-\delta}{2}\biggr)
\gamma\biggl(\frac{\delta-\beta}{2}\biggr)
\gamma\biggl(\frac{\alpha-\delta+1}{2}\biggr)
\non
&&
\hspace{2.0cm}\times
\gamma\biggl(\frac{\beta-\alpha+1}{2}\biggr)\gamma\biggl(\frac{\delta-\alpha-\beta+1}{2}\biggr)\,.
\ena
Using~\eqref{BeIntGaRSG}--\eqref{IntegrGamma}, we can
analytically derive first corrections for the structure
functions for arbitrary $\xi>0$.

\subsection{First-order corrections to the structure functions}
Applying the formulae for integrals~\eqref{BeIntGaRSG}--\eqref{IntegrGamma}
and the exact values of the CFT structure constants from Appendix A,
we now give exact expressions for the
structure functions up to the first order in $g$.

The functions $C_{\Psi\Psi}^{I}(r)$ and ${C}^{\Phi_{15}}_{\Psi\Psi}(r)$
are the simplest. (We recall that
the conformal field theory structure constants were $C_{\Psi\Psi}^{I}=1$ and
${C}^{\Phi_{15}}_{\Psi\Psi}=0$.)
Using~\eqref{BeIntGaRSG}, we find that they have the forms
\bea
\label{StruFunIFive}
&&
C_{\Psi\Psi}^{I}(r)=r^{\frac{2-\xi}{\xi+1}}+\pi g\ r^{\frac{6-\xi}{\xi+1}}\
\biggl(\frac{\gamma(\frac{\xi}{\xi+1}) \gamma^{5} (\frac{2}{\xi+1}) }
{\gamma(\frac{2\xi}{\xi+1}) \gamma(\frac{2-\xi}{\xi+1})
\gamma^2(\frac{4}{\xi+1}) }\biggr)^{\frac{1}{2}}\,,\non
&&
{C}^{\Phi_{15}}_{\Psi\Psi}(r)=-\pi g\ r^{\frac{2+7\xi}{\xi+1}}\ C_{\Psi\Psi}^{\Phi}C_{\Phi\Phi}^{\Phi_{15}}
\frac{\gamma^2(\frac{1+2\xi}{\xi+1})}{\gamma(\frac{2+4\xi}{\xi+1})}\,.
\ena

The first-order correction to the
function $C_{\Psi\Psi}^{\Phi}(r)$~\eqref{CPPPInt}
\begin{eqnarray*}
C_{\Psi\Psi}^{\Phi(1)}(r)&=&-g r^{\frac{\xi+4}{\xi+1}}\frac{\gamma(\frac{2}{\xi+1})}{\gamma^2(\frac{1}{\xi+1})}
\ \lim_{\epsilon\rightarrow 0}J(\epsilon)\,,\non
J(\epsilon)&=&J_2(-\frac{\xi}{\xi+1},-\frac{2\xi}{\xi+1},-\frac{\xi+3}{\xi+1}+2\epsilon,
\frac{\xi}{\xi+1}-\epsilon,-\frac{\xi}{\xi+1})\,
\end{eqnarray*}
is computed by applying Eq.~\eqref{IntegrGamma}. Taking the limit,
we obtain the expression
$$
\lim_{\epsilon\rightarrow 0}J(\epsilon)=
-\Bigl(\pi\frac{\xi(1-\xi)^2}{2(\xi+1)^2}\Bigr)^2
\frac{\gamma^3(\frac{1-\xi}{\xi+1})\gamma^2(\frac{\xi}{\xi+1})}
{\gamma(\frac{2-2\xi}{\xi+1})}\,.
$$
Finally, the analytic expression for the structure function
$C_{\Psi\Psi}^{\Phi}(r)$ in the first-order perturbation theory is
\be
\label{StruFunPhi}
C_{\Psi\Psi}^{\Phi}(r)=-r^{\frac{\xi}{\xi+1}}
\Bigl(\frac{\gamma(\frac{\xi}{\xi+1})\gamma(\frac{2}{\xi+1})}
{\gamma(\frac{2\xi}{\xi+1})\gamma(\frac{2-\xi}{\xi+1})}\Bigr)^{1/2}
+\pi g\ r^{\frac{\xi+4}{\xi+1}}\Bigl(\frac{\xi(1-\xi)^2}{2(\xi+1)^2}\Bigr)^2
\frac{\gamma^4(\frac{1-\xi}{\xi+1})\gamma^4(\frac{\xi}{\xi+1})}
{\gamma(\frac{2-2\xi}{\xi+1})}\,.
\end{equation}

It is convenient to rewrite structure
functions~\eqref{StruFunIFive}--\eqref{StruFunPhi}
in terms of the lightest particle mass.
Applying the mass--coupling-constant relation, we have
\begin{eqnarray}
&&
\label{FirsCorrI}
C_{\Psi\Psi}^{I}(r)=r^{\frac{2-\xi}{\xi+1}}
\left\{1-(mr)^{\frac{4}{\xi+1}}
\frac{(\xi+1)^{2}}{(\xi-1)(2\xi-1)}\left(
\frac{\gamma^{2}(\frac{\xi}{\xi+1})\gamma^{5}(\frac{2}{\xi+1})
\gamma(\frac{3\xi}{\xi+1})}
{\gamma(\frac{2\xi}{\xi+1})\gamma(\frac{2-\xi}{\xi+1})
\gamma^{2}(\frac{4}{\xi+1})}\right)^{\frac{1}{2}} \right \}\,,
\\
&&
\label{FirsCorrPhi}
C_{\Psi\Psi}^{\Phi}(r)=-r^{\frac{\xi}{\xi+1}}\Biggl\{
\left(\frac{\gamma(\frac{\xi}{\xi+1})\gamma(\frac{2}{\xi+1})}
{\gamma(\frac{2\xi}{\xi+1})
\gamma(\frac{2-\xi}{\xi+1})}\right)^{\frac{1}{2}}
\non
&&\hspace{3.0cm}
+(mr)^{\frac{4}{\xi+1}}
\frac{\xi^{2}(1-\xi)^{3}}{4(\xi+1)^{2}(2\xi-1)}\left(
\frac{\gamma^{8}(\frac{1-\xi}{1+\xi})\gamma^{9}(\frac{\xi}{\xi+1})
\gamma(\frac{3\xi}{\xi+1})}
{\gamma^{2}(\frac{2-2\xi}{\xi+1})}\right)^{\frac{1}{2}}
\Biggr\}\,,
\\
&&
\label{FirsCorrV}
C_{\Psi\Psi}^{\Phi_{15}}(r)=-r^{\frac{7\xi-2}{\xi+1}}(mr)^{\frac{4}{\xi+1}}
\frac{\xi^{2}(1-\xi)}{4(1-2\xi)(3\xi+1)^{2}}\left(
\frac{\gamma^{7}(\frac{\xi}{\xi+1})\gamma^{4}(\frac{1-\xi}{\xi+1})}
{\gamma(\frac{4\xi}{\xi+1})\gamma(\frac{2-2\xi}{\xi+1})
\gamma(\frac{2-3\xi}{\xi+1})}\right)^{\frac{1}{2}} \,.
\end{eqnarray}

We also need the structure functions for the descendent fields
in the main order. These are expressed using the corresponding
CFT structure constants of the primary fields just as~\cite{BPZ}
\begin{eqnarray}
&&C_{\Psi\Psi}^{L_{-2}\bar
L_{-2}I}(r)=\frac{\xi^2}{4(\xi+3)^2}
r^{\frac{3(\xi+2)}{\xi+1}}C_{\Psi\Psi}^{I}\,,\\
&&C_{\Psi\Psi}^{L_{-2}\bar L_{-2}\Phi}(r)=\frac{\xi^2}{4(3\xi+1)^2}
r^{\frac{5\xi+4)}{\xi+1}}C_{\Psi\Psi}^{\Phi}\,.
\end{eqnarray}

\subsection{UV expansion for the correlation function of spin operators}
Taking these formulae for structure functions, VEVs~\eqref{VEVPrimCF}
as well as the expressions
\bea
&&
\langle L_{-2}\bar{L}_{-2}I\rangle =-m^4 \Bigl(\frac{\gamma(\frac{\xi}{2})}
{\gamma(\frac{\xi+1}{2})} \Bigr)^2\,,
\non
&&\langle L_{-2}\bar{L}_{-2}\Phi\rangle =m^4
\frac{4(\xi+1)^2}{\xi^2(1-\xi)^2}
\frac{\gamma(\frac{3\xi}{2})\gamma(\frac{1-3\xi}{2})}
{\gamma(\frac{\xi}{2})\gamma(\frac{1-\xi}{2})}\ \langle \Phi\rangle  \,
\ena
for VEVs~\eqref{VEVDes} into account,
we obtain the explicit first-order UV expansion in $g$ for
the correlation function
\begin{eqnarray}
\label{AnswerShort}
\langle \Psi(r)\Psi(0)\rangle =r^{\frac{2-\xi}{\xi+1}}\left[
A(r)-B(r)Q(1-2\xi)+D(r)Q(1-4\xi) \right]\,.
\end{eqnarray}
We recall that the function $Q(\eta)$~\eqref{Qfactor} for the
fields $\Phi_{1,2s+1}$ is given by the Eqn.~\eqref{QgammaI}\ -\
\eqref{QgammaII}. For the fields  $\Phi_{13}$ and $\Phi_{15}$ the
respective expressions in~\eqref{AnswerShort} turn out to be
\begin{eqnarray}
&&
Q(1-2\xi)=\frac{(1-\xi)(1-2\xi)}{2(\xi+1)}\:\gamma\left(\frac{\xi}{2}\right)
   \gamma\left(\frac{1-\xi}{2}\right)
   \left[\gamma\left(\frac{1}{\xi+1}\right)
   \gamma\left(\frac{1-2\xi}{\xi+1}\right)\right]^{\frac{1}{2}}\,,\\
&&\nonumber\\
&&\nonumber\\
&&
Q(1-4\xi)=\frac{(1-\xi)(1-2\xi)(1-3\xi)(1-4\xi)}{4(\xi+1)^2}\:\gamma\left(\frac{\xi}{2}\right)
    \gamma\left(\frac{3\xi}{2}\right)\nonumber\\
&&
\hspace{1.8cm}\times\gamma\left(\frac{1-\xi}{2}\right)\gamma\left(\frac{1-2\xi}{2}\right)
\gamma\left(\frac{1-3\xi}{2}\right)
   \left[
\gamma\left(\frac{1}{\xi+1}\right)
   {\gamma\left(\frac{1-4\xi}{\xi+1}\right)}
\right]^{\frac{1}{2}}.
\end{eqnarray}
The $r$-dependent functions $A(r)$, $B(r)$, and $D(r)$
in~\eqref{AnswerShort} have the forms
\begin{eqnarray}
&& A(r)= 1-\frac{\xi^{2}}{4(\xi+3)^{2}}
\frac{\gamma^{2}(\frac{\xi}{2})}{\gamma^{2}(\frac{1+\xi}{2})}
(mr)^{4}
\nonumber\\
&&\hspace{1.7cm}
-\ \frac{(\xi+1)^{2}}{(1-\xi)(1-2\xi)}\left(
\frac{\gamma^{2}(\frac{\xi}{\xi+1})\gamma^{5}(\frac{2}{\xi+1})
\gamma(\frac{3\xi}{\xi+1})}
{\gamma(\frac{2\xi}{\xi+1})\gamma(\frac{2-\xi}{\xi+1})
\gamma^{2}(\frac{4}{\xi+1})}\right)^{\frac{1}{2}}
(mr)^{\frac{4}{\xi+1}}\,,
\\
\nonumber\\
&&B(r)=
\left\{
\left(\frac{\gamma(\frac{\xi}{\xi+1})\gamma(\frac{2}{\xi+1})}
{\gamma(\frac{2\xi}{\xi+1})\gamma(\frac{2-\xi}{\xi+1})}
\right)^{\frac{1}{2}}
\left[1+\frac{(\xi+1)^{2}}{(\xi-1)^{2}(3\xi+1)^{2}}
\frac{\gamma(\frac{3\xi}{2})\gamma(\frac{1-3\xi}{2})}
{\gamma(\frac{\xi}{2})\gamma(\frac{1-\xi}{2})}
(mr)^{4}\right] \right.\nonumber\\
&&
\hspace{1.0cm} \left.
+\frac{\xi^{2}(1-\xi)^{3}}{4(\xi+1)^{2}(1-2\xi)}\left(
\frac{\gamma^{8}(\frac{1-\xi}{1+\xi})\gamma^{9}(\frac{\xi}{\xi+1})
\gamma(\frac{3\xi}{\xi+1})}
{\gamma^{2}(\frac{2-2\xi}{\xi+1})}\right)^{\frac{1}{2}}
(mr)^{\frac{4}{\xi+1}}\right\} (mr)^{\frac{2(\xi-1)}{\xi+1}}\,,
\\
&& D(r)=- 
\frac{\xi^{2}(1-\xi)}{4(1-2\xi)(3\xi+1)^{2}}\left(
\frac{\gamma^{7}(\frac{\xi}{\xi+1})\gamma^{4}(\frac{1-\xi}{\xi+1})}
{\gamma(\frac{4\xi}{\xi+1})\gamma(\frac{2-2\xi}{\xi+1})
\gamma(\frac{2-3\xi}{\xi+1})}\right)^{\frac{1}{2}}
(mr)^{\frac{8\xi}{\xi+1}}\,.
\end{eqnarray}

\subsection{Model ${{\cal M}}_{2,7}$ as an example}
The first model in the series ${{\cal M}}_{2,5}$ is identified with the
Lee-Yang model \cite{Cardy} with the central charge of Virasoro algebra $c=-\frac{22}{5}$.
There are only two primary fields in the theory: the unit
field $I:=\Phi_{11}= \Phi_{14}$ with the conformal dimension zero and
the spin field $\Phi_{12}=\Phi_{13}$ with the dimension $-\frac{1}{5}$.
This example was analyzed in detail in~\cite{AlZamLY,FFLZZ} .

The next example is the model ${{\cal M}}_{2,7}$ \cite{Gehlen} with the central charge
\be
\label{CenChaACFT}
c=-\frac{68}{7}\,.
\end{equation}
In this case, there are three primaries: the unit field $I:=\Phi_{11}=
\Phi_{16}$, the spin field $\Psi:= \Phi_{12}=\Phi_{15}$, and the energy field
$\Phi:= \Phi_{13}=\Phi_{14}$ with the respective conformal dimensions
\bea
\label{SpectSevCFT}
\Delta_I\ =\hspace{0.3cm}0\,, \quad \Delta_\Psi=-\frac{2}{7}\,, \quad\Delta_\Phi=-\frac{3}{7}\,.
\ena

The first-order corrections in $g$ to the structure functions are given by
expressions~\eqref{FirsCorrI}--\eqref{FirsCorrV}
with $\xi=\frac{2}{5}$.

Providing an analytic continuation of the Lukyanov--Zamolodchikov
solution, we have the following data for the VEVs of
operators giving the leading contribution:
\begin{eqnarray}
&&m_1^{\frac{6}{7}}\lan\Phi_{13}\ran=2.269550689\,,\nonumber
\\\nonumber
&&m_1^{\frac{4}{7}}\lan\Phi_{15}\ran =-2.325136 \ i \,
\end{eqnarray}
for primary fields and
\bea
&&m_1^{\frac{8}{7}} C^{L_{-2}\bar{L}_{-2}I}_{\Psi\Psi}(r)
\lan L_{-2}\bar{L}_{-2}I\ran=
\non
&&
\hspace{3.0cm}=
-0.0005899419474(m_1r)^{\frac{36}{7}}\,,\non
&&m_1^{\frac{8}{7}} C^{L_{-2}\bar{L}_{-2}\Phi}_{\Psi\Psi}(r)
\lan L_{-2}\bar{L}_{-2}\Phi(0)\ran=
\non
&&
\hspace{3.0cm}
=0.01123584810(m_1r)^{\frac{30}{7}}\nonumber
\ena
for the first nontrivial descendants.
Taking into account that
the mass coupling constant for $\xi=2/5$ is
\be
g=-0.04053795542378225 m_1^{20/7}\,
\end{equation}
and collecting all data together, we obtain the short-distance expansion
up to $r^6$:
\bea
&&m_1^{\frac{8}{7}}\lan\Psi(x)\Psi(0)\ran\sim -5.83 (m_1r)^{\frac{2}{7}}
+(m_1r)^{\frac{8}{7}}-0.265(m_1r)^{22/7}+0.233(m_1r)^{24/7}\non
&&\hspace{3.1cm} -
0.045(m_1r)^4+0.011(m_1r)^{\frac{30}{7}}\non
&&\hspace{3.1cm} -0.0006(m_1r)^{\frac{36}{7}}+
O(r^{6})\,.
\ena
The plots for this function are shown in Figs.~\ref{corrfun1}--\ref{corrfun3}.
It can be clearly seen that the contributions from the descendent fields
$\lan L_{-2}\bar{L}_{-2}\Phi\ran$ and $\lan L_{-2}\bar{L}_{-2}I\ran$
are essential.

\section{Long-distance expansion}
Massive models~\eqref{ActiRSG} allow an alternative description
as scattering theories with factorized S matrices~\cite{SmirnovQG,LeClair}.
In this section, we use the form-factor approach to
study the long-distance behaviors of the spin correlation functions.
The structure of ground states of the model as well as
the spectrum of particles depend on the numbers
$(p,p')$~\cite{ABF,SmirnovQG,RSOS}.
There are two kinks interpolating between $p-1$ different
vacuums and $[(p'-p)/p]$ bound states.

For simplicity, we restrict our attention to the case of
nonunitary models ${{\cal M}}_{2,2n+1}$
having only $n$ primary
fields $\Phi_{1,k}=\Phi_{1,2n-k+1},
(k=1,\ldots, n)$. The corresponding
perturbed models are the simplest integrable two-dimensional models
in the sense that the scattering matrices are diagonal.
The multiparticle
form factors of the fields $\Psi,\Phi,\ldots$ in these theories
do not contain complicated contour integrals originating from
solitonic form factors~\cite{Smirnov}. Hereafter in this section,
we use the symbol $\xi$ only for
$$
\xi=\frac{2}{2n-1}\,.
$$
We note that for studying a general case, one must also take
kink contributions into account.

\subsection{Perturbed ${{\cal M}}_{2,2n+1}$ models as scattering theories.}
We briefly recall the basic facts about the scattering theories under
consideration (see, e.g,~\cite{Koubek} for a detailed review).
In ${{\cal M}}_{2,2n+1}$ models perturbed by the field $\Phi$,
there are $j=1,\ldots, n-1$ scalar self-conjugate particles with the masses
\be
\label{MasParFF}
{
m_j:=m_1\frac{\sin(\frac{\pi \xi j}{2})}{\sin(\frac{\pi \xi}{2} )}\,. 
}
\end{equation}
Here, $m_1$ is a mass of the lightest particle ``1".
The standard momentum parameterization is
given in terms of rapidities $\beta$ as
\bea
\label{MomeFF}
&&p^{0}_a=m_a\cosh \beta\ , \non
&&p^{1}_a=m_a\sinh \beta\ .
\ena

The S matrix for these models originates from the breathers S matrix of the
corresponding sine-Gordon model~\cite{SmirnovQG,FreClaMel}:
\be
\label{SMatGeFF}
{
S_{ab}(\beta)=f_{\frac{|a-b|}{2}}(\beta) f_{\frac{a+b}{2}}(\beta)
\prod_{s=1}^{min(ab)-1}\bigl(f_{\frac{|a-b|+2s}{2}}(\beta) \bigr)^2\,.
}
\end{equation}
Here, the basic building block $f_{a}(\beta)$ is defined through the
meromorphic functions
\be
\label{FunEfFF}
{
f_{a}(\beta)=\frac{\tanh\frac{1}{2}(\beta+i \pi a\xi)}{\tanh\frac{1}{2}(\beta-i\pi a\xi)}\,.
}
\end{equation}
In the physical strip $0<\mathrm{Im}\,\beta<\pi$, the
scattering matrices satisfy the
unitarity and crossing symmetry conditions given respectively as
\begin{eqnarray*}
S_{ab}(\beta)S_{ba}(-\beta)=1\,,\hspace{1.0cm}
S_{ab}(\beta)=S_{ab}(i\pi-\beta)\,,
\end{eqnarray*}
as well as the additional bound-state condition described as follows.
If $S_{ab}(\beta)$ has a simple pole $\beta_{ab}=iu_{ab}^c$ in the
direct channel, then the particle ``c" of mass
$$
m_c^2=m_a^2+m_b^2+2m_am_b \cos(u_{ab}^c )
$$
is a bound state of ``a" and ``b".
The bound state
$|{{B}}_c(\beta)\ran$ with $\beta=\beta_a+\beta_b$ is defined as
the projection on the pole of the two-particle
state $|B_a(\beta_a)B_b(\beta_b)\ran$ at the relative
rapidity $\beta_{ab}=iu_{ab}^c$.
Then the S matrices for the composite particle can be found via
the bootstrap equations
$$
S_{cd}(\beta)=S_{ad}\bigl(\beta+i(\pi-{u}_{ac}^b)\bigr)S_{bd}\bigl(\beta-i(\pi-{u}_{bc}^a)\bigr)\,.
$$
In particular, the total scattering matrix $S_{ab}$~\eqref{SMatGeFF}
can be reconstructed starting from the S matrix
$S_{11}(\beta)=f_{1}(\beta)$ for the lightest particle
assuming that the bootstrap tree for our massive models is closed under the
fusions~\cite{FreClaMel}
\begin{eqnarray*}
&&a_i\times a_j\longrightarrow a_{i+j} \quad \hbox{or}\quad a_i\times a_j\longrightarrow a_{2n-1-i-j}\ , \non
&&
a_i\times a_j\longrightarrow a_{i-j}\ .
\end{eqnarray*}
The rule in the first line above is to choose the variant
where a final particle $i+j$ or $2n-1-i-j$ is in the
range $(1,\ldots, n-1)$.



\subsection{Form-factor approach}
As we already mentioned, the correlation functions
admit a form-factor expansion that is very useful for studying the
infrared behaviors. For example, for spin fields, the corresponding
spectral decomposition is
\bea
\label{CorFunFF}
&&\lan\Psi(x)\Psi(0)\ran=\sum_{n=0}^{\infty}\frac{1}{n!}\int
\frac{d \beta_1}{2\pi}\cdots \frac{d \beta_n}{2\pi}
\lan 0|\Psi(x)|\beta_1,\ldots,\beta_n\ran_{a_1\cdots a_n} \times\non
&&\hspace{4.2cm}\times{}_{a_1,\ldots,a_n}\lan \beta_1,\ldots,\beta_n|\Psi(0)|0\ran=
\non
&&\hspace{2.2cm}=\sum_{n=0}^{\infty}\frac{1}{n!}\sum_{\{a_j\}}\int
\frac{d \beta_1}{2\pi}\cdots \frac{d \beta_n}{2\pi}
e^{-r\sum_j m_{a_j}\cosh \beta_j}\non
&&\hspace{4.2cm}\times
F_{a_n\cdots
a_1}(\beta_n,\ldots, \beta_1) F_{a_1\cdots
a_n}(\beta_1,\ldots, \beta_n)\ ,
\ena
where we introduce the matrix elements of the local operators
in the basis of the asymptotic states
\be
\label{FormFGenFF}
{
F_{a_1\cdots a_n}(\beta_1,\ldots, \beta_n)=
\lan 0|\Psi(0)|\beta_1,\ldots,\beta_n\ran_{a_1\cdots a_n}\ .
}
\end{equation}
Form factors~\eqref{FormFGenFF} are defined in this approach as a set
of functions satisfying
Smirnov's axioms~\cite{Smirnov} (also see~\cite{Karowski}). We
recall them for completeness.
The first two axioms are the Watson equations
\bea
\label{WatsonFF}
&&F_{a_{1}\cdots a_{j}a_{j+1}\dots}
(\beta_1,\ldots,\beta_{i},\beta_{i+1},\cdots,\beta_n)=\non
&&\quad \quad \quad \quad =S_{a_{j}a_{j+1}}(\beta_{j}-\beta_{j+1})
F_{a_{1}\cdots a_{j+1} a_{j}\cdots a_n}
(\beta_1,\ldots,\beta_{j+1},\beta_{j},\ldots,\beta_n)\ ,\non
&&F_{ a_{1} a_{2}\cdots a_{n}}(\beta_{1}+2\pi i,\beta_{2},\cdots,\beta_{n})=\non
&&\quad \quad \quad \quad =
F_{ a_{2}\cdots a_{n} a_{1}}(\beta_{2},\cdots,\beta_{n},\beta_{1})\ .
\ena
The requirement for relativistic invariance implies that for the
local operator with spin $s$, one has
\be
\label{RelatInvFF}
{
F_{a_{1}\cdots a_{n}}
(\beta_1+\Lambda,\cdots,\beta_n+\Lambda)=e^{s\Lambda}
F_{a_{1}\cdots a_{n}}
(\beta_1,\cdots,\beta_n)}\,.
\end{equation}
The ultraviolet bound condition requires
that there exist a finite constant $t(j,n)<\infty$ such that
\be
\label{GrowthFF}
{
F_{a_{1}\cdots a_{n}}
(\beta_1+\Lambda,\cdots,\beta_j+\Lambda,\beta_{j+1},\ldots,\beta_n)=
O(\exp e^{t(j,n)|\Lambda|})\ , \quad \Lambda\longrightarrow \infty\ .
}
\end{equation}
In addition, there are two pole conditions. The first is
a kinematic pole condition. In our case for self-conjugate particles,
it looks like
\bea
\label{KinPolFF}
&&-i\lim_{\beta'\rightarrow\beta}(\beta^{'}-\beta)
F_{a a a_{1}\dots a_n}(\beta^{'}+i\pi,\beta,\beta_{1},\ldots,\beta_n)=\non
&&\quad \quad \quad \quad =
(1-\prod_{j=2}^{n}S_{aa_{j}}(\beta-\beta_{j}))
F_{a_{1}\cdots a_{n}}(\beta_{1},\ldots,\beta_{1})\ .
\ena
The second pole condition is the equation for bound states.
Let $a_i\times a_j\longrightarrow a_k$. Then
\bea
\label{BoundStaFF}
&&-i\lim_{\beta^{'}\rightarrow\beta}(\beta^{'}-\beta)
F_{a_{1}\dots a_{i}a_{j}\cdots a_{n}}
(\beta_1,\ldots,\beta^{'}+i\bigl(\pi-u_{a_{i}a_{k}}^{a_{j}}\bigr)
,\beta-i\bigl(\pi-u_{a_{j}a_{k}}^{a_{i}}\bigr),\dots)=\non
&&\quad \quad \quad \quad =
\Gamma_{a_{i}a_{j}}^{a_{k}}
F_{a_{1}\cdots a_{k}\cdots a_{n}}(\beta_1,\ldots,\beta,\ldots,
\beta_n)\,,
\ena
where the three particle on-shell vertex is defined as\footnote{We note that
the perturbed minimal models ${{\cal M}}_{2,2n+1}$
violate the one-particle unitarity since some of its couplings
are purely imaginary.}
$$
-i \hbox{Res} \ S_{ab}(\beta)\bigl|_{\beta=i u_{ab}^c}=(\Gamma_{ab}^{c})^2\ .
$$
Smirnov~\cite{Smirnov} proved
that the set of functions~\eqref{FormFGenFF} satisfying
conditions~\eqref{WatsonFF}--\eqref{BoundStaFF}
defines a matrix element of a local operator
in the scattering theory with the given S matrix.

\subsection{First form factors}
The solutions of locality axioms~\eqref{WatsonFF}--\eqref{BoundStaFF}
for the primaries
in perturbed ${{\cal M}}_{2,2n+1}$ models have already been discussed in the
literature.
The form factors of the fields
$\Psi$ and $\Phi$ were found in~\cite{SmirnovQG} via the quantum group
reduction procedure from the sine-Gordon model.
The case of the scaling Lee--Yang model
${{\cal M}}_{2,5}$ was directly elaborated in~\cite{AlZamLY}. The general
expressions for the primaries $\Phi_{1k}$ appeared in~\cite{Koubek} based
on the form factors for exponential operators in the
sinh-Gordon model~\cite{MussarI}. But for our aim of studying the
correlation functions that are in agreement with~\eqref{TwoPRSG},
we must complete these
prescriptions by choosing a proper overall factor determining
normalization~\eqref{Normal} of local operators. This was essentially
provided in~\cite{LuSine}.\footnote{Another advantage of
Lukyanov's bosonization rules is that they can be easily extended
to include the kink sector for a
perturbed minimal CFT with general $(p,p')$.}
Indeed, after the
quantum group reduction~\cite{AlZamMassMu, FLZZ},
the expressions for the first breather form factors of the
exponential operators in the sine-Gordon model
can be essentially treated as the following particle ``1"
form factors of the primaries
in the reduced model:\footnote{We note that the
same result can be obtained using the exact realization of the
form factors in the RSOS model and taking the scaling limit~\cite{VEVAlg}.}
\begin{eqnarray}
\label{SerLukOFF}
&&\lan 0|\Phi_{1k}|0\ran=\lan\Phi_{1k}\ran\,,
\\
\label{SerLukIFF}
&& \lan 0|\Phi_{1k}|\beta\ran =i \ C
\frac{\sin{\Bigl((k-1)\pi \xi/2}\Bigr)}{
\sin(\pi\xi)}\ \lan\Phi_{1k}\ran\,,
\\
\label{SerLukIIFF}
&&\lan 0|\Phi_{1k}|\beta_2,\beta_1\ran =i^2 C^2
\frac{\sin^2{\Bigl((k-1)\pi \xi/2}\Bigr)}{
\sin^2(\pi\xi)}\
R(\beta_1-\beta_2)\ \lan\Phi_{1k}\ran \,,
\\
\label{SerLukIIIFF}
&&\lan 0|\Phi_{1k}|\beta_3, \beta_2,\beta_1\ran =i^3 C^3
\frac{\sin{\Bigl((k-1)\pi \xi/2}\Bigr)}{
\sin(\pi\xi)}
\prod_{s<j}R(\beta_s-\beta_j) \nonumber\\
&&\hspace{0.5cm}\times\Bigl\{
\frac{\sin^2{\Bigl((k-1)\pi \xi/2}\Bigr)}{
\sin^2(\pi\xi)}
+
\frac{1}{\prod_{s<j}2\cosh(\frac{\beta_s-\beta_j}{2})}
\Bigr\}
\ \lan\Phi_{1k}\ran \,,
\ena
and so on. Here, the function $R(\beta)$ determining
the rapidity-dependent part of
the two-particle form factor is given explicitly as
\be
\label{ScaFunFF}
R(\beta) = \exp\Bigl\{
4\int_0^\infty \frac{dt}{t} \frac{\sinh t\sinh \xi t \sinh
(\xi+1)t}{\sinh^2 2t} \cosh 2(1-\frac{i}{\pi}\beta) t \Biggr\} \,.
\end{equation}
%
We note that it has a property that is very useful
for checking the locality axioms at low particle levels (while one of
the easiest ways to prove the validity of the general formulae is to use the
free-field realization):
\bea
\label{PropRFuFF}
&&R(\beta)R(\beta\pm i\pi)
=\frac{\sinh(\beta)}{\sinh(\beta)\mp i\sin(\pi \xi)} \,,
\non
&&
\prod_{k=0}^{n-2}R\left(\beta+\frac{i\pi\xi}{2}(2k+1)\right)
R\left(\beta-\frac{i\pi\xi}{2}(2k+1)\right)
=\frac{\cosh(\beta)-\cos(\pi\xi)}{\cosh(\beta)+1}R(\beta)\,.
\ena

The $\beta$-independent constant $C$ fixing the normalization
of the first particles is
\bea
C^2
=8\cos^2(\frac{\pi\xi}{2})
\sin(\frac{\pi\xi}{2})
\exp\Bigl(-\int_0^{\pi \xi}\frac{d t}{\pi}\frac{t}{\sin t}\Bigr)\,.
\ena

\subsection{Correlation functions at long distances}
One can see that the leading terms for the correlation function already
come from the zero- and one-particle form factors.
For example, even the first two-term expression
\bea
\label{OnePartFirCF}
&&\lan\Phi_{1k}(x)\Phi_{1k}(0)\ran=\non
&&\qquad=\lan\Phi_{1k}\ran^2\Bigl\{1-\frac{2}{\pi}\frac{\sin^2(\frac{\pi}{2}(k-1)\xi)}{\sin(\frac{\pi}{2}\xi)
}\
\exp\Bigl(-\int_0^{\pi \xi}\frac{d t}{\pi}\frac{t}{\sin t}\Bigr)
K_0(m_1r)\Bigr\}+\cdots\,,
\ena
where $K_0(x)$ is a Macdonald function,
turns out to be a good approximation for the correlation
functions at long distances.

Of course, for more accurate calculations, one must take
higher form factors with more particles into account, including those with
particles $a_j$ with
$j>1$. The latter can be easily derived directly through the
bound-state axiom~\eqref{BoundStaFF}.\footnote{The additional requirements
for the consistency of formulae~\eqref{SerLukOFF}--\eqref{SerLukIIIFF},
etc., appearing from the bound-state condition are automatically satisfied
for the operators $\Phi_{1,k}$ as was argued in~\cite{Koubek}. For
form factors with a small number of particles, this can be
verified straightforwardly.}
It is clear from~\eqref{CorFunFF} that at long
distances the contribution
of an $n$-particle form factor with
particles $a_1,\ldots,a_n$ and masses
$m_{a_1},\ldots,m_{a_n}$ to the correlation function
is essentially approximated by the quantity $\exp(-r\sum_j m_{a_j})$.
In practice, the form factors with a higher number of particles
$a_1$ and
particles of high mass become important only at small distances.
It is useful to order the form factors according to the value of this
quantity. Ror example, for the perturbed ${{\cal M}}_{2,7}$ model
that is considered further,
the leading terms thus come from $F_1$, $F_2$, $F_{11}$, $F_{21}$,
$F_{111}$, etc., in agreement with mass relations~\eqref{MasParFF}
$m_1<m_2<2m_{1}<m_1+m_2<3m_1$, etc.

\section{Numerical results}
The form-factor and conformal perturbation expansions work
well at long and short distances respectively.
In this section, we collect the numerical data demonstrating matching
between these two at $mr\sim 1$
for the case of scaling ${{\cal M}}_{2,7}$ models as an example.
Substituting $\xi=\frac{2}{5}$ for the ${{\cal M}}_{2,7}$ case, we obtain
the expression
$$
\lan\Phi_{12}|B_1(\beta)\ran\lan B_1(\beta)
|\Phi_{12}\ran=-\lan\Phi_{12}\ran^2\cdot 0.7574894945\,.
$$
Taking into account the numerical value for the VEV
\be
m_1^{\frac{4}{7}}\lan\Phi_{12}\ran= -2.325598436 \ i\,,
\end{equation}
we find that the leading term in the long-distance expansion
is given by the formula
\be
\label{FirsLar}
m_1^{\frac{8}{7}}\lan\Phi_{12}(x)\Phi_{12}(0)\ran\sim
-5.408408086\ \Bigl(1-0.2411163946 \cdot K_0(rm_1)\Bigr)\,.
\end{equation}

Comparison of the short- and long-distance expansions for the two-point
correlation function $\langle \Psi (x) \Psi (0)\rangle$ is shown in
Figs.~\ref{corrfun1}--\ref{corrfun3}.

\begin{figure}[!htb]
\vspace{2mm}
\begin{center}
\epsfxsize=10.0cm \epsffile{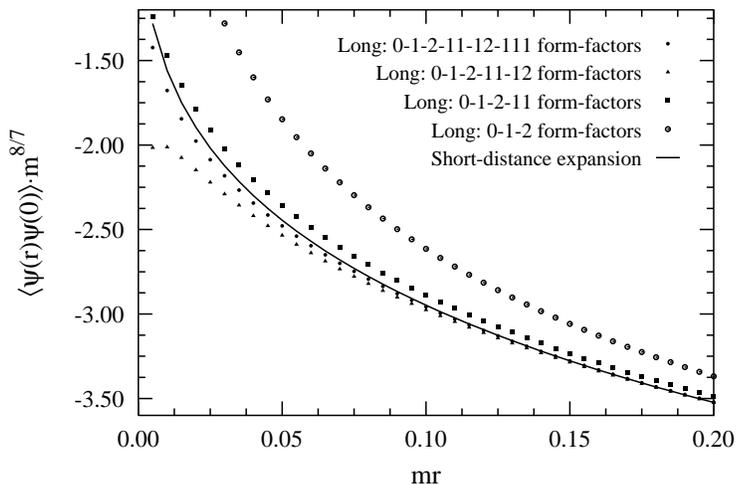}
\end{center}
\caption{Form factor contributions at small distances}
\label{corrfun2}
\end{figure}

The numerically calculated data is presented in Table~\ref{tbl1}.
The difference between the short- and long-distance expansions
indicates the self-consistency of the zeroth-order perturbation theory and
the form-factor expansion up to two particles.

\begin{figure}[!htb]
\vspace{2mm}
\begin{center}
\epsfxsize=14.0cm \epsffile{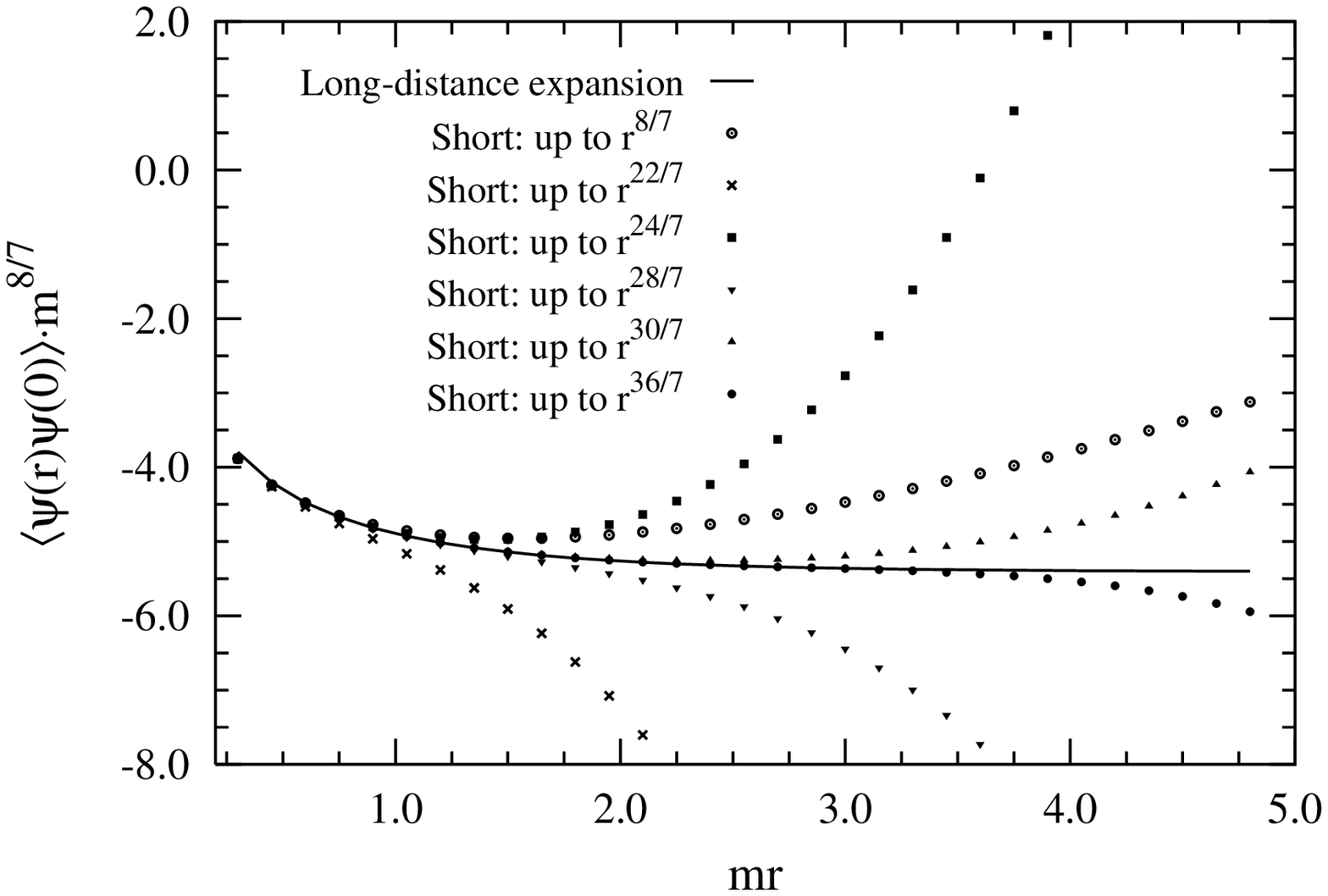}
\end{center}
\caption{Behaviors of different terms of the short distance expansion}
\label{corrfun3}
\end{figure}

\begin{center}
\begin{table}
\begin{tabular}{|c|c|c|c|c|c|c|}
\hline
&\multicolumn{3}{c|}{\it Short-distance expansion}&\multicolumn{3}{c|}{\it Long-distance expansion} \\
\hline
$mr$&up to $r^{8/7}$&up to $r^{28/7}$&up to $r^{36/7}$&up to $F_{11}$&up to $F_{12}$&up to $F_{111}$\\
\hline
0.02&-1.89536&-1.89536&-1.89536&-1.78911&-2.14822&-1.97677\\
0.04&-2.29916&-2.29917&-2.29917&-2.20386&-2.41955&-2.34353\\
0.06&-2.56977&-2.56979&-2.56979&-2.48822&-2.63930&-2.59599\\
0.08&-2.77773&-2.77778&-2.77778&-2.70767&-2.82117&-2.79351\\
0.10&-2.94806&-2.94817&-2.94817&-2.88736&-2.97622&-2.95732\\
0.12&-3.09288&-3.09306&-3.09306&-3.03986&-3.11138&-3.09787\\
0.14&-3.21905&-3.21935&-3.21934&-3.17245&-3.23117&-3.22120\\
0.16&-3.33092&-3.33135&-3.33135&-3.28975&-3.33870&-3.33115\\
0.18&-3.43139&-3.43200&-3.43199&-3.39489&-3.43619&-3.43036\\
0.20&-3.52254&-3.52336&-3.52335&-3.49009&-3.52528&-3.52071\\
0.22&-3.60588&-3.60696&-3.60694&-3.57702&-3.60725&-3.60361\\
0.24&-3.68258&-3.68397&-3.68395&-3.65692&-3.68306&-3.68014\\
0.26&-3.75354&-3.75530&-3.75526&-3.73077&-3.75352&-3.75114\\
0.28&-3.81950&-3.82167&-3.82162&-3.79937&-3.81926&-3.81732\\
0.30&-3.88103&-3.88367&-3.88361&-3.86334&-3.88081&-3.87921\\
0.35&-4.01850&-4.02259&-4.02247&-4.00626&-4.01911&-4.01810\\
0.40&-4.13682&-4.14280&-4.14259&-4.12947&-4.13911&-4.13845\\
0.50&-4.33032&-4.34153&-4.34097&-4.33216&-4.33782&-4.33752\\
0.70&-4.60061&-4.62933&-4.62699&-4.62274&-4.62492&-4.62484\\
0.90&-4.77130&-4.82907&-4.82226&-4.82018&-4.82110&-4.82108\\
1.00&-4.83076&-4.90821&-4.89757&-4.89617&-4.89679&-4.89677\\
1.20&-4.91088&-5.03975&-5.01671&-5.01633&-5.01661&-5.01661\\
1.40&-4.95019&-5.14911&-5.10491&-5.10540&-5.10554&-5.10554\\
1.60&-4.95765&-5.24871&-5.17110&-5.17252&-5.17258&-5.17258\\
1.80&-4.93933&-5.34862&-5.22123&-5.22367&-5.22371&-5.22371\\
2.00&-4.89960&-5.45779&-5.25949&-5.26302&-5.26304&-5.26304\\
2.20&-4.84170&-5.58470&-5.28902&-5.29349&-5.29349&-5.29349\\
2.40&-4.76811&-5.73780&-5.31231&-5.31721&-5.31721&-5.31721\\
2.60&-4.68080&-5.92578&-5.33150&-5.33577&-5.33577&-5.33577\\
2.80&-4.58133&-6.15775&-5.34855&-5.35034&-5.35034&-5.35034\\
3.00&-4.47099&-6.44339&-5.36541&-5.36181&-5.36181&-5.36181\\
3.20&-4.35086&-6.79303&-5.38416&-5.37087&-5.37087&-5.37087\\
3.40&-4.22182&-7.21775&-5.40703&-5.37805&-5.37805&-5.37805\\
3.60&-4.08465&-7.72941&-5.43656&-5.38374&-5.38374&-5.38374\\
\hline
\end{tabular}
\caption{}
\label{tbl1}
\end{table}
\end{center}

\section{Conclusion}
We have demonstrated that there is a region in the mass scale
where the long- and short-distance expansions for the correlation
functions of the spin operators
match each other. We can therefore conclude that the combination
of the two approaches allows
reconstructing the correlation functions at all scales.

This also serves as a good consistency check for the
correctness of both methods as well as the VEVs of local operators.

\section*{Acknowledgments}
We thank B.~Feigin, V.~Fateev, A.~Zamolodchikov, and S.~Lukyanov
for the useful discussions and also W.~Everett for editorial
assistance. This work was supported in part by the grants INTAS
00-00055,
 RFBR-01-02-16686, RFBR-02-01-01015, LS-2044.20032, NATO grant
 PST.CLG.979008 and by Program of RAS ``Elementary particles."
Y.P. was also supported by Russian Science Support Foundation.
%

When the paper was completed we learned from V.~Fateev that
the quantities $J_2(a,b,$ $d,e,c)$ in Eqn. (\ref{DoublNumARSG}) at
the special values of parameters as in the l.h.s. of Eq.
(\ref{IntegrGamma}) can be transformed to the class of integrals
studied in \cite{Fateev},  where analytical expressions in terms
of gamma functions had already been proposed. We are grateful to 
V.~Fateev for informing us about this.

\section*{Appendix}
\appendix
\section{Structure constants and conformal blocks}
For completeness, we here give
a derivation of structure constants for
the subalgebra of conformal fields $\Phi_{1k}$
and relevant conformal blocks that are necessary for building the
corresponding correlation functions in conformal
field theory.
To reproduce these known results~\cite{DotFat},
we use the method of functional equations~\cite{ZZCFT}.

We consider the four-point conformal block with the chiral
$\Phi_{12}$ operator
\be
\label{FourCFT}
G(z)=\langle \Phi_{{12}}(z)\Phi_{\Delta_{1}}(0)\Phi_{\Delta_{2}}(1)\Phi_{\Delta_{3}}(\infty)\rangle \ ,
\end{equation}
where $\Phi_{\Delta_j}(z)$ denotes chiral parts of the primary field
$\Phi_{1n_j}(z,\bar{z})$ $(k=1,...,p'-1)$
with the conformal dimensions $\Delta_{1,n_j}$.
This holomorphic function satisfies the null vector equation~\cite{BPZ}
\be
\label{NullCFT}
\mu \frac{d^2}{d^2z} G(z)+\bigl(\frac{1}{z}+\frac{1}
{z-1}\bigr)\frac{d}{dz} G(z)
+\bigl(\frac{\Delta_1+\Delta_2+\delta-\Delta_3}{z(z-1)} -
\frac{\Delta_1}{z^2}-\frac{\Delta_2}{(z-1)^2} \bigr) G(z)=0
\,
\end{equation}
with $\mu=\frac{3}{2(2\Delta_{12}+1)}$.
The solutions of the equation and their monodromy properties can be easily
found by noting that it turns out to be the Riemann equation
\bea
\label{RiemCFT}
&&\frac{d^2u}{d^2z}+\Bigl(\frac{1-\alpha-\alpha'}{z}+\frac{1-\gamma-\gamma'}{z-1}\Bigr)\frac{du}{dz}
\non
&&\hspace{0.8cm}
+\Bigl(\frac{\alpha\alpha'}{z^2} +\frac{\gamma\gamma'}
{(z-1)^2}+\frac{\beta\beta'-\alpha\alpha'-
\gamma\gamma'}{z(z-1)} \Bigr)u=0\,, \\
&&
\alpha+\alpha'+\beta+\beta'+\gamma+\gamma'=1\nonumber
\ena
with the parameters
\begin{eqnarray*}
&&\alpha=\frac{n_1-1}{2}\rho\,,\hspace{1.5cm} \alpha'=1-\frac{n_1+1}{2}\rho\,,\non
&&\beta=\frac{2-n_3}{2}\rho\,,\hspace{1.5cm} \beta'=\frac{n_3+2}{2}\rho-1\,,\non
&&\gamma=\frac{n_2-1}{2}\rho\,,\hspace{1.5cm} \gamma'=1-\frac{n_2+1}{2}\rho\,.
\end{eqnarray*}
Here, the parameter $\rho$ is related to the labels
$(p,p')$ of the minimal model as $\rho=p/p'$.

The equation~\eqref{RiemCFT} has two linearly independent solutions
having power-law behaviors at $z=0$.
These correspond to two conformal blocks in the S-channel,
\bea
\label{SChanCFT}
&&S_1(z)=z^{\alpha}(1-z)^{\gamma}
\
{}_2F_1
\Bigl(\begin{matrix}&\alpha+\beta+\gamma,&\alpha+\beta'+\gamma\cr
        &1+\alpha-\alpha'&
\end{matrix}
\Bigl| z\Bigr)
\ ,\non
&&S_2(z)=z^{\alpha'}(1-z)^{\gamma}
\
{}_2F_1
\Bigl(\begin{matrix}&\alpha'+\beta+\gamma,&\alpha'+\beta'+\gamma\cr
        &1+\alpha'-\alpha&
\end{matrix}
\Bigl| z\Bigr)
\ .
\ena
Another set of linear independent solutions having power law form at $z=1$
corresponds to T-channel conformal blocks:
\bea
\label{TChanCFT}
&&T_1(z)(z)=z^{\alpha}(1-z)^{\gamma}
\
{}_2F_1
\Bigl(\begin{matrix}&\alpha+\beta+\gamma,&\alpha+\beta'+\gamma\cr
        &1+\gamma-\gamma'&
\end{matrix}
\Bigl| 1-z \Bigr)
\ ,\non
&&T_2(z)(z)=z^{\alpha}(1-z)^{\gamma'}
\
{}_2F_1
\Bigl(\begin{matrix}&\alpha+\beta+\gamma',&\alpha+\beta'+\gamma'\cr
        &1+\gamma'-\gamma&
\end{matrix}
\Bigl| 1-z \Bigr)
\ .
\ena
Since second-order differential equation~\eqref{RiemCFT}
has two linear independent solutions, $S_1$ and $S_2$ are expressed in terms
of $T_1$ and $T_2$ as
\bea
\label{MonodCFT}
&S_1(z)=AT_1(z)+BT_2(z)\ ,\non
&S_2(z)=CT_1(z)+DT_2(z)\ .
\ena
Using the relations between solutions of Riemann equations
(or just using the formulae for the analytic continuation
for the hypergeometric functions), we obtain
the expressions for the $z$-independent coefficients
\begin{eqnarray*}
\label{MonoMatCFT}
&&A=\frac{\Gamma(1+\alpha-\alpha')\Gamma(\gamma'-\gamma)}{\Gamma(\alpha+\beta'+\gamma')
\Gamma(\alpha+\beta+\gamma')}\ , \non
&&B=\frac{\Gamma(1+\alpha-\alpha')\Gamma(\gamma-\gamma')}{\Gamma(\alpha+\beta+\gamma)
\Gamma(\alpha+\beta'+\gamma)}\ , \non
&&C=\frac{\Gamma(1+\alpha'-\alpha)\Gamma(\gamma'-\gamma)}{\Gamma(\alpha'+\beta'+\gamma')
\Gamma(\alpha'+\beta+\gamma')}\ , \non
&&D=\frac{\Gamma(1+\alpha'-\alpha)}{\Gamma(\alpha'+\beta+\gamma)}
\frac{\Gamma(\gamma-\gamma')}{\Gamma(\alpha'+\beta'+\gamma)}\,.
\end{eqnarray*}

In what follows, we need the ratio $AB/CD$.
It can be expressed in terms of the $\gamma(x)$ function as
\begin{eqnarray*}
\frac{AB}{CD}=-\frac{\gamma(1+\alpha-\alpha')}{\gamma(1+\alpha'-\alpha)}
\frac{\gamma(\alpha'+\beta+\gamma')\gamma(\alpha'+\beta+\gamma)\gamma(\alpha'+\beta'+\gamma)}
{\gamma(\alpha+\beta+\gamma)}\,.
\end{eqnarray*}
For the conformal blocks in model ${{\cal M}}_{p,p'}$,
the following expression holds for the operators $\Phi_{1,n_j}$:
\begin{eqnarray*}
\frac{AB}{CD}=-\frac{\gamma\bigl(n_1\rho\bigr)}
{\gamma\bigl(2-n_1\rho\bigr)}
\frac{\gamma\bigl(2-(n_1+n_2+n_3)\frac{\rho}{2}\bigr)}
{\gamma\bigl((n_1-n_2+n_3)\frac{\rho}{2}\bigr)}
\frac{\gamma\bigl((-n_1+n_2+n_3)\frac{\rho}{2}\bigr)}
{\gamma\bigl((n_1+n_2-n_3)\frac{\rho}{2}\bigr)}\,.
\end{eqnarray*}
\begin{figure}[!htb]
\vspace{2mm}
\begin{center}
\epsfxsize=10.0cm \epsffile{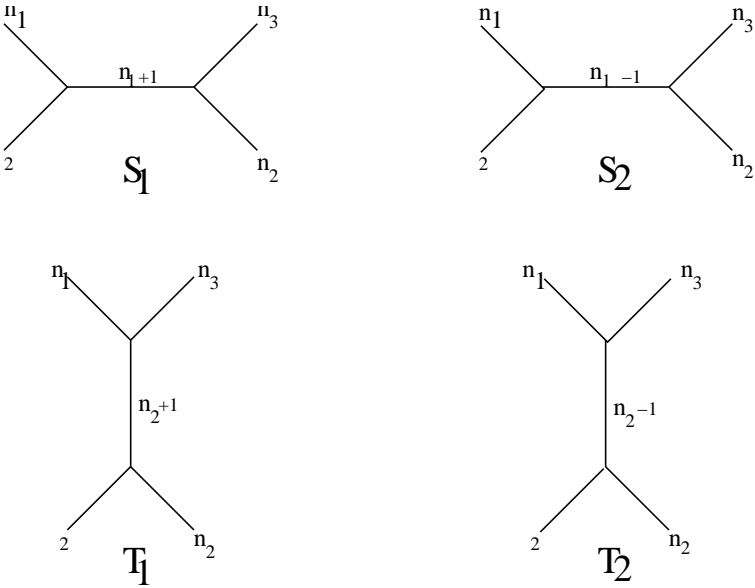}
\end{center}
\caption{Conformal blocks.
}
\label{crossingSym}
\end{figure}
The crossing symmetry condition~\cite{BPZ} (see fig. \ref{crossingSym}) leads to the
equation for the correlation functions
\begin{eqnarray*}
&&C(2,n_1,n_1+1)C(n_1+1,n_2,n_3)|S_1|^2
+C(2,n_1,n_1-1)C(n_1-1,n_2,n_3)|S_2|^2=\non
&&=C(2,n_2,n_2+1)C(n_2+1,n_1,n_3)|T_1|^2+
C(2,n_2,n_2-1)C(n_2-1,n_1,n_3)|T_2|^2 \,,
\end{eqnarray*}
where $C(n_1,n_2,n_3)$ denotes the structure constants $C_{n_1n_2}^{n_3}$
of the operator algebra of fields $\Phi_{1,n}$.
In particular, this gives
$$
C(2,n_1,n_1+1)C(n_1+1,n_2,n_3)AB
+C(2,n_1,n_1-1)C(n_1-1,n_2,n_3)CD=0 \ .
$$

We consider the case $n_3=n_1=n$ and $n_2=2$. If $A_0,B_0,\ldots$
are the corresponding elements of the monodromy matrix, then we find
$$
C^2(2,n,n+1)A_0B_0+C^2(2,n,n-1)C_0D_0=0 \ .
$$
Therefore,
$$
\Bigl[\frac{C(2,n,n+1)}{C(2,n,n-1)}\Bigr]^2=
-\frac{C_0D_0}{A_0B_0}\ .
$$
We can now write the general difference equation for the
structure constants in the form
$$
\frac{C(n_1+1,n_2,n_3)}{C(n_1-1,n_2,n_3)}=
-\Bigl[-\frac{A_0B_0}{C_0D_0}\Bigr]^{\frac{1}{2}}\times
\frac{CD}{AB}\ .
$$
Rewriting everything via the variables $n_j$, we obtain
the functional equation
for structure constants in the $\Phi_{1,n}$ subalgebra of ${{\cal M}}_{p,p'}$
\bea
\label{InterCCFT}
&&\frac{C(n_1+2,n_2,n_3)}{C(n_1,n_2,n_3)}
=\Bigl[
\frac{\gamma\bigl(2-(n_1+1)\rho\bigr)}
{\gamma\bigl(n_1\rho\bigr)}
\frac{\gamma\bigl(2-(n_1+2)\rho\bigr)}
{\gamma\bigl((n_1+1)\rho\bigr)}
\Bigr]^{\frac{1}{2}}\times\non
&&\times\frac{\gamma\bigl((n_1-n_2+n_3+1)\frac{\rho}{2}\bigr)}
{\gamma\bigl((2-n_1+n_2+n_3+1)\frac{\rho}{2}\bigr)}
\frac{\gamma\bigl((n_1+n_2-n_3+1)\frac{\rho}{2}\bigr)}
{\gamma\bigl((-n_1+n_2+n_3-1)\frac{\rho}{2}\bigr)}
\ .
\ena
We seek the solution in the ansatz
$$
C(n_1,n_2,n_3)=N(n_1)\prod_{k=1}^{n}\frac{\gamma\bigl(k\rho\bigr)}{\gamma\bigl((n_1-k)\rho\bigr)
\gamma\bigl((n_2-k)\rho\bigr)\gamma\bigl(2-(n_3+k)\rho\bigr)}\ ,
$$
where
$$
n=(n_1+n_2-n_3-1)/2 \ .
$$
Substituting this expression in~\eqref{InterCCFT}, we obtain the
constraint for $N(n_1)$
$$
\frac{N(n_1+2)}
{N(n_1)}=\Biggl(\gamma\bigl(n_1\rho\bigr)\gamma\bigl((n_1+1)\rho\bigr)
\gamma\bigl(2-(n_1+1)\rho\bigr)\gamma\bigl(2-(n_1+2)\rho \bigr)\Biggr)^{\frac{1}{2}}\ ,
$$
which can be easily resolved as
\be
\label{NOneCFT}
N(n)=\Biggl(\prod_{k=1}^{n-1}\gamma\bigl(k\rho\bigr)\gamma\bigl(2-(k+1)\rho\bigr)\Biggr)^{\frac{1}{2}}\ .
\end{equation}
Taking the symmetry $\Delta_i\leftarrow \rightarrow \Delta_j$ into account,
we find the expression for the structure constants
\be
\label{StruCoCFT}
C(n_1,n_2,n_3)=
\frac{N(n_1)N(n_2)}{N(n_3)}
\prod_{k=1}^{n}\frac{\gamma\bigl(k\rho\bigr)}{\gamma\bigl((n_1-k)\rho\bigr)
\gamma\bigl((n_2-k)\rho\bigr)\gamma\bigl(2-(n_3+k)\rho\bigr)}\ .
\end{equation}
We briefly comment on the admissible triples $(n_1,n_2,n_3)$. The procedure
used in the derivation assumes that the field $\Phi_{1,n_3}$
must be related to $\Phi_{1,n_1}$ and $\Phi_{1,n_2}$
by fusion rules~\cite{BPZ}. Otherwise, the proposed
structure constants should be reconstructed by symmetry.

\section{Integrals for the first corrections}
The integrals from the correlation functions that appear
at the first order correction can be represented in the
form of double integrals over a plane
because of the equation \cite{DotFat,DotPP,GuiMag}
\begin{eqnarray}
\label{BB}
&&
\hspace{-0.35cm}\pi\frac{\gamma(a+1)\gamma(c+1)}{\gamma(a+c+2)}
\int \dd^2x |x|^{2(a+d+c+1)}|1-x|^{2e}
\ \Bigl|\ {}_2F_1
\Bigl(\begin{matrix}&a+1,&-b\cr
        &a+c+2&
\end{matrix}
\Bigl| x\Bigr)\Bigr|^2
\nonumber\\
&& +
\pi\frac{\gamma(b+1)\gamma(a+c+1)}{\gamma(a+b+c+2)}
\int \dd^2x |x|^{2d}|1-x|^{2e}
\ \Bigl|\ {}_2F_1
\Bigl(\begin{matrix}&-a-b-c-1,&-c\cr
        &-a-c&
\end{matrix}
\Bigl| x\Bigr)\Bigr|^2
\nonumber\\
&& ={\int \dd^2x \int \dd^2y
    |x|^{2a} |1-x|^{2b} |y|^{2d} |1-y|^{2e} |x-y|^{2c} }
\nonumber\,.
\end{eqnarray}

For practical computations it is convenient
to decompose  this integral into a sum of holomorphic and antiholomorphic parts
using the way proposed in works \cite{Lipatov, DotFat}
(see \cite{ConstFlum,GuiMag} for alternative methods).
Starting with the integral
\begin{equation}
    J=\int\int
d^2x d^2y\:|x|^{2a} |1-x|^{2b} |y|^{2d} |1-y|^{2e}
    |x-y|^{2c}\,,
\end{equation}
performing Wick rotation
\begin{eqnarray*}
x_2=i x_0 (1-2 i \epsilon) \,,
\qquad y_2=i y_0 (1-2 i \epsilon)\,,
\end{eqnarray*}
and introducing the new variables
\begin{equation*}
    \begin{alignedat}{2}
       &x=x_1+x_0\,, &\qquad
       &{\bar x}=x_1-x_0\,,\\
       &y=y_1+y_0\,, &\qquad
       &{\bar y}=y_1-y_0\,,
    \end{alignedat}
\end{equation*}
we can easily obtain
\begin{equation}\label{int}
 \begin{split}
    J=&-\frac{1}{4}\iint  dx dy\: x^a (x-1)^b y^d (y-1)^e (x-y)^c\\
&\times\iint d{\bar x}d{\bar y}\: ({\bar x}+i \epsilon(x-{\bar
x}))^a ({\bar x}-1+i \epsilon (x-{\bar x}))^b
({\bar y}+i \epsilon(y-{\bar y}))^d \\
&\qquad\qquad\times({\bar y}-1+i \epsilon (y-{\bar y}))^e ({\bar
x}-{\bar y}+i \epsilon ((x-{\bar x})-(y-{\bar y})))^c \,.
  \end{split}
\end{equation}
Integral \eqref{int} has the branch points
\begin{equation}
 \begin{aligned}
    &{\bar x}=-i\epsilon x\,,\\
    &{\bar x}=1-i\epsilon (x-1)\,,\\
    &{\bar x}={\bar y}-i\epsilon(x-y)\,,
 \end{aligned}
 \qquad
 \begin{aligned}
  &{\bar y}=-i\epsilon y\,,  \\
  &\bar{y}=1-i\epsilon(y-1)\,.
 \end{aligned}
\end{equation}
By deforming integration contours is easy to demonstrate that non-trivial
contributions to \eqref{int} come only from two domains of integration,
namely,
\be
   \{0<x<1\,,\quad 0<y<x\}\,
  \qquad  \hbox{and}\qquad
   \{0<x<1\,,\
\quad x<y<1\}\,.
\en
One observes that the integral $J$ is represented as sum
\begin{equation}
\label{PartIntI}
    J=I(a,b,d,e,c)+I(d,e,a,b,c)\,,
\end{equation}
where
\begin{multline}
I(a,b,d,e,c)=-\frac{1}{4} \int\limits_0^1 dx \int\limits_0^x dy \:x^a (x-1)^b y^d (y-1)^e (x-y)^c\times\\
\times\int\limits_{C_1} d{\bar x} \int\limits_{C_2} d{\bar y}\:
{\bar x}^a ({\bar x}-1)^b {\bar y}^d ({\bar y}-1)^e ({\bar
x}-{\bar y})^c \,,
\nonumber
\end{multline}
\begin{multline}
I(d,e,a,b,c)=-\frac{1}{4}\int\limits_0^1 d x \int\limits_x^1 d y x^a\: (x-1)^b y^d (y-1)^e (x-y)^c\times\\
      \times\int\limits_{C_4} d{\bar y}\int\limits_{C_3} d{\bar x}\:
         {\bar x}^a ({\bar x}-1)^b {\bar y}^d ({\bar y}-1)^e ({\bar x}-{\bar y})^c \,,
\nonumber
\end{multline}
We introduce notations
$I(a,b,d,e,c), \ I(d,e,a,b,c)$ for the integrals above since
that two integrals are related by
transform $a\rightarrow d$, $b\rightarrow e$
as it can be easily checked by change of variables.
\begin{figure}[!htb]
\vspace{2mm}
\begin{center}
\epsfxsize=12.0cm \epsffile{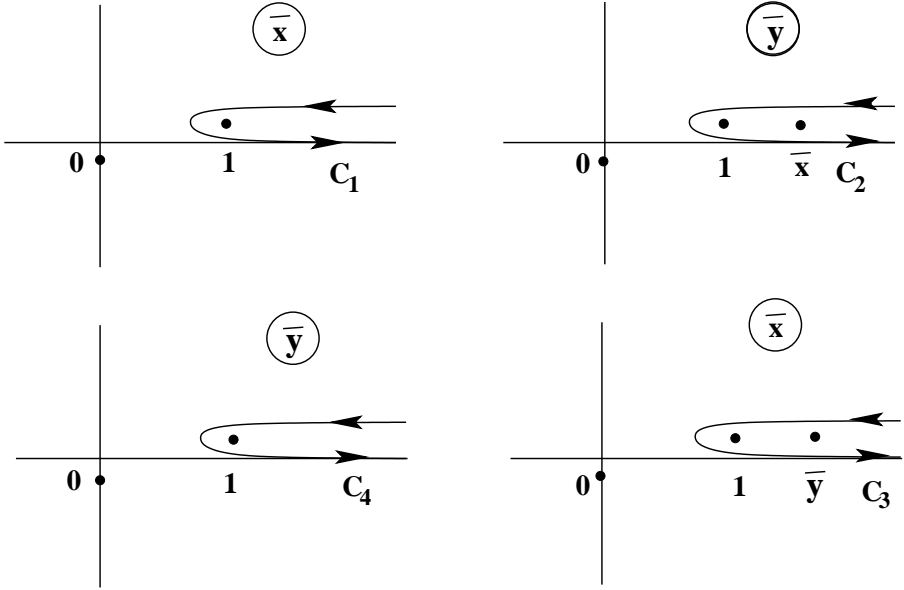}
\end{center}
\caption{The contours $C_1,C_2,C_3,C_4$.} \label{fig:conturs}
\end{figure}
The contours of integrations are shown in
Fig.~\ref{fig:conturs}.
Let us now transform the
contour integrals $I(a,b,d,e,c)$ into
standard type integrals.
This is easily provided by applying the technique
of analytical continuations and manipulation with
the contours described in \cite{DotFat}.
We obtain the following expression
\begin{equation}
\label{PartIntIII}
    I(a,b,d,e,c)=\sin \pi b K(a,b,d,e,c)[\sin \pi e L(a,b,d,e,c) +
     \sin \pi(e+c)L(d,e,a,b,c)]\,
            ,
\end{equation}
where
\begin{equation}
    \begin{aligned}
    &K= \int\limits_0^1 d x \int\limits_0^1 d y\: x^{a+d+c+1} (1-x)^b y^d (1-y)^c
    (1-xy)^e\,,\\
    &L=\int\limits_0^1 d y \int\limits_0^1 d x\: x^{-a-b-c-2} (1-x)^c y^{-a-b-d-e-c-3}
    (1-y)^e(1-xy)^b\,.
    \end{aligned}
\end{equation}
Now one uses standard formulae for the integral representations of the
hypegeometric functions
\begin{eqnarray*}
&&\int\limits_0^1 d y\: y^p (1-y)^q (1-xy)^r = B(p+1,q+1)
\
_2F_1
\Bigl(\begin{matrix}&-r,&p+1\cr
        &p+q+2&
\end{matrix}
\Bigl| x\Bigr)
\,,
\non
&&\int\limits_0^1 x^s (1-x)^t
\
_2F_1
\Bigl(\begin{matrix}a,&b\cr
        c&
\end{matrix}
\Bigl| x\Bigr)
d x  =
 B(s+1,t+1) \
_3F_2
\Bigl(\begin{matrix}&a,& b,& s+1\cr
        &s+t+2,&c &
\end{matrix}
\Bigl| 1\Bigr)\,,
\end{eqnarray*}
to find that the integrals $K$ and $L$ admit another representation in terms of the
higher hypergeometric functions. Here $B(x,y)=\Gamma(x)\Gamma(y)/\Gamma(x+y)$ is a beta function.
Namely,
\bea
&&K=B(d+1,c+1) B(a+d+c+2,b+1)
\nonumber\\
&&\hspace{0.3cm}\times \
 _3F_2\Bigl(
\begin{matrix}&-e,& a + d + c + 2,& d + 1\cr
        &a + d + c + 3 + b,& d + 2 + c&
\end{matrix}
\Bigl| 1\Bigr)\,,
\non
&&
\label{HighHyp}\\
&&
L=B(-a-b-d-e-c-2,e+1) B(-a-b-c-1,c+1)
\nonumber \\
&&\hspace{0.3cm}\times
\
_3F_2
\Bigl(\begin{matrix}&-b,& -a-b-d-e-c-2,& -a-b-c-1\cr
        &-a-b-d-c-1,& -a-b&
\end{matrix}
\Bigl| 1\Bigr)\,.
\nonumber
\ena

Equations (\ref{PartIntI}),(\ref{PartIntIII})
and (\ref{HighHyp})
give a form convenient for numerical studies.
We note that the functions $_3F_2$ at unity satisfy specific
identities \cite{Prud}. This
simplifies the answer and leads to (\ref{IntegrGamma}).


\begin{thebibliography}{99}
\bibitem{AlZamLY}
{Zamolodchikov Al.~B.: Two-point correlation function
in scaling Lee-Yang model. Nucl. Phys. {\bf B348}, 619-641 (1991)}

\bibitem{BPZ}
{Belavin, A.~A., Polyakov, A.~M. and Zamolodchikov A.~B.:
Infinite conformal symmetry in two-dimensional quantum field
theory. Nucl. Phys. {\bf B241}, 333-380 (1984) }


\bibitem{McCoySato}
{Wu, T.~T., McCoy, B.~M., Tracy, C.~A. and Barouch, E.:
Spin-spin correlation functions for the
two-dimensional Ising model: exact
theory in the scaling region
{ Phys. Rev.} {\bf B13},
316-374 (1976) \\
Sato, M., Jimbo, M. and Miwa, T.: { Publ. Res. Inst. Intern. Math. Sci.
} {\bf 14}, 223 (1978); {\bf 15}, 201, 557, 871 (1979); {\bf 16}, 531 (1980)
}



\bibitem{Zam}
{Zamolodchikov A.~B. Integrable field theory from
conformal field theory. Adv. Stud. in Pure Math. {\bf 19}, 641-674
(1989) }

\bibitem{ABF}
{Andrews, G., Baxter, R. and Forrester, J.: Eight-vertex
SOS model and generalized Rogers-Ramanujan identities. J. Statist.
Phys. {\bf 35}, 193-266 (1984)}


\bibitem{Karowski}
{ Karowski, M.  and Weisz, P.: Exact form factors in
{($1+1$)}-dimensional field theoretic models with soliton
behavior. Nucl. Phys. {\bf B139}, 455--476 (1978) }

\bibitem{Smirnov}
{Smirnov, F.~A.: Form-factors in completely
integrable models of quantum field theory. Singapore: World
Scientific (1992) }



\bibitem{AlZamMassMu}
{ Zamolodchikov Al.~B.: Mass Scale In The Sine-Gordon
Model And Its Reductions. Int.J.Mod.Phys., {\bf A10}, 1125  (1995)
}





\bibitem{DotFat}
{ Dotsenko, Vl. S. and Fateev, V. A.: Conformal
algebra and multi-point correlation functions in 2d statistical
models. Nucl. Phys. {\bf B240} \ [{\bf  FS12}], 312-348 (1984)
\\
Dotsenko, Vl. S. and  Fateev, V. A.: Four-point correlation
functions and the operator algebra in 2d conformal invariant
theories with central charge $c\le1$. Nucl. Phys. {\bf B251}\
[{\bf FS13}] 691-734 (1985)}


\bibitem{SmirnovQG}
{ Smirnov F.~A.: Reductions of quantum Sine-Gordon
model as perturbations of minimal models of conformal field
theory. Nucl. Phys. {\bf B337}, 156-180 (1990)}

\bibitem{LeClair}
{LeClair, A.:
Restricted sine-Gordon theory and the minimal conformal series
Phys. Lett. {\bf B230}, 103-107 (1989) }

\bibitem{RSOS}
{Bazhanov, V.V. and Reshetikhin N.Yu.: Scattering amplitudes
in off-critical models and RSOS integrable models.
{ Prog. Theor. Phys. Supplement.} {\bf 102},  301-318 (1990)}


\bibitem{GuiMag}
{Guida, R., and Magnoli, N.: All order IR finite
expansion for short distance behavior of massless theories
perturbed by a relevant operator. Nucl. Phys. {\bf B471}, 361-388
(1996)}

\bibitem{Cardy}
{Cardy, J.~L.: Conformal invariance and the Lee-Yang edge singularity
 in two dimensions.
{ Phys. Rev. Lett.} {\bf 54}, 1354-1356  (1985) }

\bibitem{Gehlen}
{von Gehlen, G.: Non-hermitian triticality in Blume-Capel model
  with imaginary field.
 { Int. J. Mod. Phys.}, {\bf B8}, 3507, (1994)}

\bibitem{Lipatov}
{Lipatov, L.~N.: Calculation of the Gell-Mann-Low function in a scalar
field theory with strong nonlinearity. { JETP}, {\bf 44}, 1055 (1976);
{ Zh. Eks. Teor. Fiz.},{\bf 71}, 2010-2024 (1976)}


\bibitem{Mussar}
{Mussardo, G.: Off-critical statistical models:
factorizable scattering theories and bootstrap programm. Phys.
Rep. {\bf 218}, 215-379 (1992)}



\bibitem{LZ}
{Lukyanov, S. and Zamolodchikov A.: Exact expectation
values of local fields in quantum sine-Gordon model. Nucl. Phys.
{\bf B 493}, 571-587 (1997)}

\bibitem{FLZZ}
{Fateev, V.,  Lukyanov, S. Zamolodchikov A. and
Zamolodchikov Al.: Expectation values of local fields in
Bullough-Dodd model and integrable perturbed conformal field
theories. { Nucl. Phys.}
{\bf B516}, 652-674 (1998)}

\bibitem{FFLZZ}
{Fateev, V., Fradkin, D.,   Lukyanov, S. Zamolodchikov A. and
Zamolodchikov Al.: Expectation values of descendents fields in the
sine-Gordon model. { Nucl. Phys.} {\bf B540} 587-609 (1999)
[ArXiv:hep-th/9807236] }

\bibitem{FateevIII}
{ Baseilhac P.,  Fa\-teev V.~A.: Ex\-pectation values
of lo\-cal fields for a two-pa\-ra\-me\-ter family of integrable models
and related perturbed conformal field theories. Nucl. Phys. {\bf
B532}, 567-587 (1998) }

\bibitem{FateevI}
{Fateev, V.~A.: Normalization factors in conformal
field theory and their applications. Mod. Phys. Lett. {\bf A15}
259-270 (2000)}

\bibitem{FateevII}
{
Ahn, C., Fateev, V.~A., Kim, C., Rim, C., Yang, B.:
Reflection Amplitudes of ADE Toda Theories and
Thermodynamic Bethe Ansatz.
Nucl. Phys. {\bf B565}, 611-628 (2000)
}

\bibitem{ConstFlum}
{Constantinescu, F., and Flume, R.: Perturbation Theory around
two-dimensional critical systems through horomorphic
decomposition. J.Phys. {\bf A23}, 2971 (1990)}

\bibitem{DotPP}
{Dotsenko, V., Picco M. and Pujoi, P.: Renormalization group calculation
of correlation functions for the 2D random bound Ising and Potts models.
Nucl. Phys. {\bf B455}, 701-723 (1995).
}

\bibitem{FreClaMel}
{
Freund, P.G., Klassen, T.R., Melzer, E.:
S Matrices for perturbations of certain conformal field theories
{ Phys. Lett.} {\bf B229}, 243 (1989).
}

\bibitem{Koubek}
{Koubek, A.: Form-factor bootstrap and the operator
content of perturbed minimal models.
Nucl. Phys. {\bf B428}, 655-680 (1994)
}





\bibitem{MussarI}{Koubek, A. and Mussardo, G.: On the operator content
of the sine-Gordon model. Phys. Lett. {\bf B311}, 193-201 (1993) }

\bibitem{LuSine}
{ Lukyanov S.: Form-factors of exponential fields in
the {sine-Gordon} model. Mod. Phys. Lett. {\bf A12}, 2543-2550 (1997)}

\bibitem{VEVAlg}
{Pugai, Y.: On vertex operators and the normalization of form-factors. In
{\it  Statistical field theories, 57-66, eds. A.~Cappelli, G.~Mussardo,
Kluwer Academic Press, 2002.}}


\bibitem{ABFII}
{Jimbo, M., Konno, H., Odake, S., Pugai, Y. and Shiraishi, J.:
Free field construction for ABF models in the regime II.  { J.
Stat. Phys.} {\bf 102}, 883-921 (2001) [arXiv:math.qa/0001071].
}

\bibitem{MagnoliII}
Caselle, M., Grinza, P. and Magnoli, N.:
Short distance behaviour of correlators in the
2D Ising model in a magnetic field. [arXiv:hep-th/9909065]



\bibitem{LukDoy}
{Doyon, B., Lukyanov, S.: Fermion Schwinger's function for the
SU(2) Thirring model. [arXiv:hep-th/0203135] }


\bibitem{CFTPerturb}
Fioravanti, D., Mussardo G., and Simon, P.: Universal Amplitude
Ratios of The Renormalization Group: Two-Dimensional Tricritical
Ising Model, { Phys.\ Rev.}\ E {\bf 63}, 016103 (2001)
[arXiv:cond-mat/0008216]


\bibitem{ABFIII}{ Lukyanov S., Pugai Y.: Multipoint Local Height
Probabilities in the Integrable RSOS Model, Nucl. Phys. {\bf B
473}, 631-658, (1996) }


\bibitem{ZZCFT}{Zamolodchikov A.~B., Zamolodchikov Al.~B.:
Conformal field theory and critical phenomena in
two-dimensional systems. { Physics reviews} v.{10},
269-433 (1989). (Ed.) I.M. Khalatnikov,
London UK: Harwood 1989 (Soviet scientific reviews.
Section A.10.4)}


\bibitem{Prud}
{Prudnikov A.~P., Brychkov Yu.~A. and Marichev O.~I.:  Integrals and Series,Vol.3,
 Gordon Breach Science Publishers, 1990 }



\bibitem{Fateev} Fateev V.~A.: unpublished.





\end{thebibliography}
\end{document}